# CRYPTOPLANET UPDATE

Robert L. Kurucz

Harvard-Smithsonian Center for Astrophysics

February 15, 2007

Abstract

We have had several talks recently reviewing 11 years of exoplanet discoveries through radial velocity variations, or from transits, or from microlensing. More than 200 exoplanets have been found, including some around pulsars that we do not discuss here.

My physical definition for a planet is a roughly spherical, self-gravitating body more massive than $10^{26}$ g formed from the leftover material in a protostellar disk after the protostar forms. Radiation from the protostar pushes the inner wall of the disk outward. The material agglomerates and forms planets in radial sequence. The outer planets are formed slowly by classical dynamical mechanisms acting in the snow zone. Planets have dense cores because of agglomeration. Planets bound around other stars were formed and behave like the planets in our solar system.

Not one of the exoplanets discovered thus far is a planet. They are cryptoplanets formed from matter ejected by protostars. When protostars have excessive infall at high latitudes, they partially balance angular momentum through outflow at the equator as they spin up. The ejected matter is trapped in the magnetic torus formed between the star and the disk, like a tokamak. The tokamak eventually reconnects and magnetic compression forms self-gravitating remnants trapped and compressed by a closed spherical magnetic field, spheromaks. Cooled spheromaks are cryptoplanets. They orbit near the star. They can merge with each other or fall into the star or be ejected. They can grow by accreting gas. They have a low density core and abundances characteristic of the protostar. Their masses, radii, densities, and orbits are random, and are inconsistent with the parameters for planets. They tend to have lower density than planets.

More than 20 disks around protostars have been observed and are claimed to be pre-planetary. In fact they are post-planetary. Disks are not observable until the central Jupiter-orbit-size region has been cleared out and the inner planets have been formed out through "Jupiter".

# CRYPTOPLANET UPDATE

## ROBERT L. KURUCZ

## HARVARD-SMITHSONIAN
## CENTER FOR ASTROPHYSICS

## FEBRUARY 15, 2007

# CRYPTOPLANET UPDATE

**CONSERVATION OF ANGULAR MOMENTUM**

**CONVECTION**

**UNIFORMITARIANISM**

**FORMATION OF PLANETS**

**FORMATION OF CRYPTOPLANETS**

**CRYPTOPLANET CHARACTERISTICS**

# CONSERVATION OF ANGULAR MOMENTUM

SHU (1982)

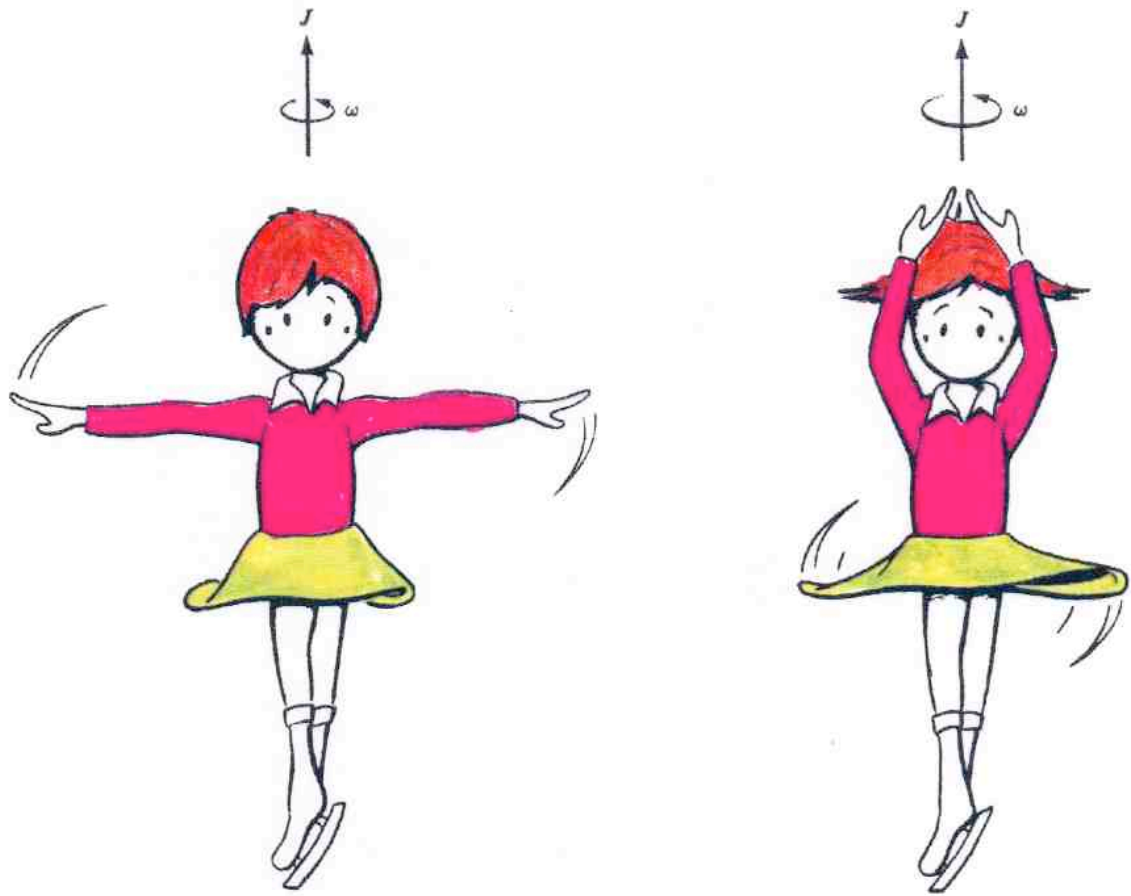

**Figure 3.23.** The spin-up of an ice-skater. When the twirling skater has her arms out, her moment of inertia $I$ is large and her rate of spin $\omega$ is small for a given angular momentum $J = I\omega$. When she pulls her arms in, her moment of inertia $I$ decreases, and the conservation of angular momentum implies that her rate of spin $\omega$ increases.

**HOWEVER, THIS CAPTION DOES NOT EXPLAIN THE FIGURE**

**MASSES:**
  **SKATER 50 KG, HAIR 50G, SKIRT 500G**

**1/1000 OF MASS MOVES OUTWARD IN HAIR**
**1/100  OF MASS MOVES OUTWARD IN SKIRT**

**ADD 1 KG LEAD DRAPERY WEIGHTS TO HEM OF SKIRT.  NOW 3/100 OF MASS.
ENOUGH TO REDUCE SPINUP**

**ADD TO CAPTION:
A ROTATING BODY TRIES TO MINIMIZE ITS SPINUP BY MAXIMIZING ITS MOMENT OF INERTIA**

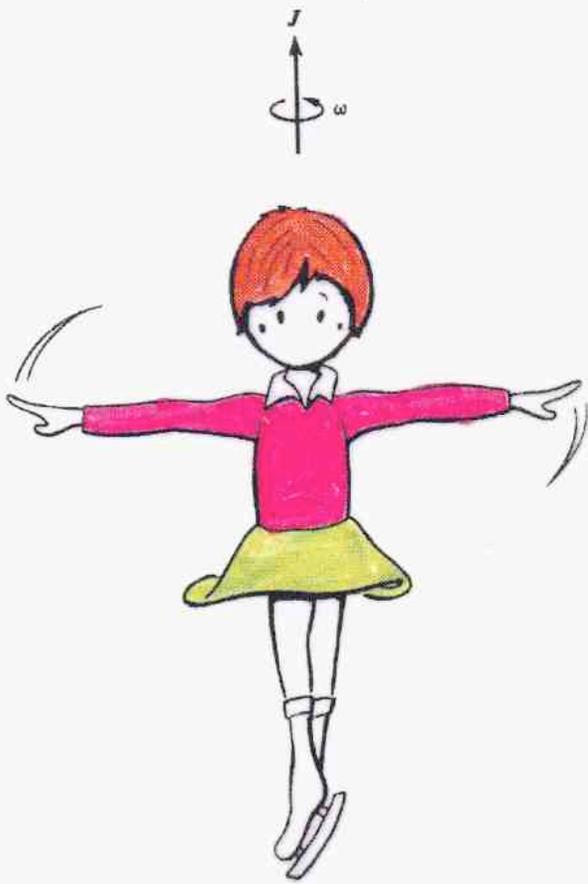
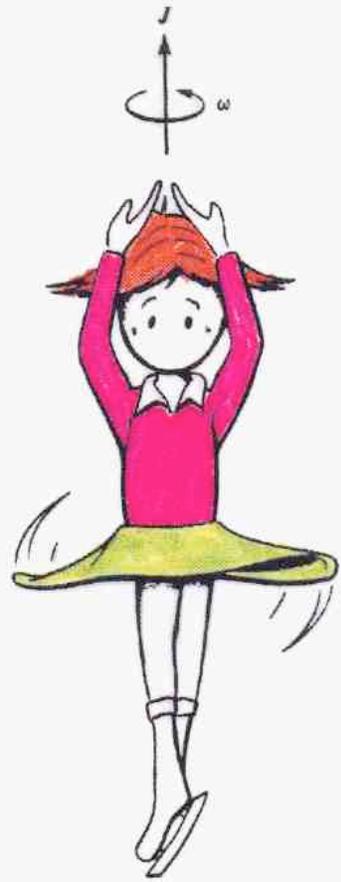
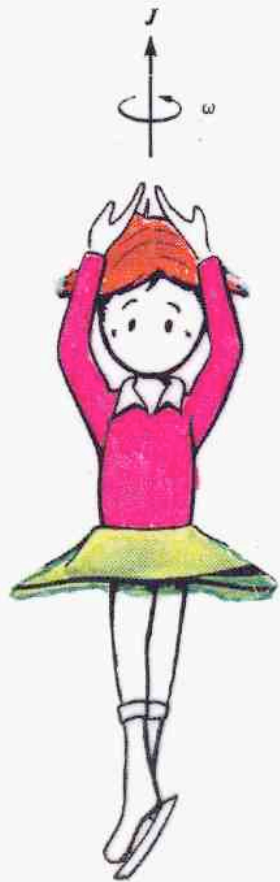

# CONVECTION

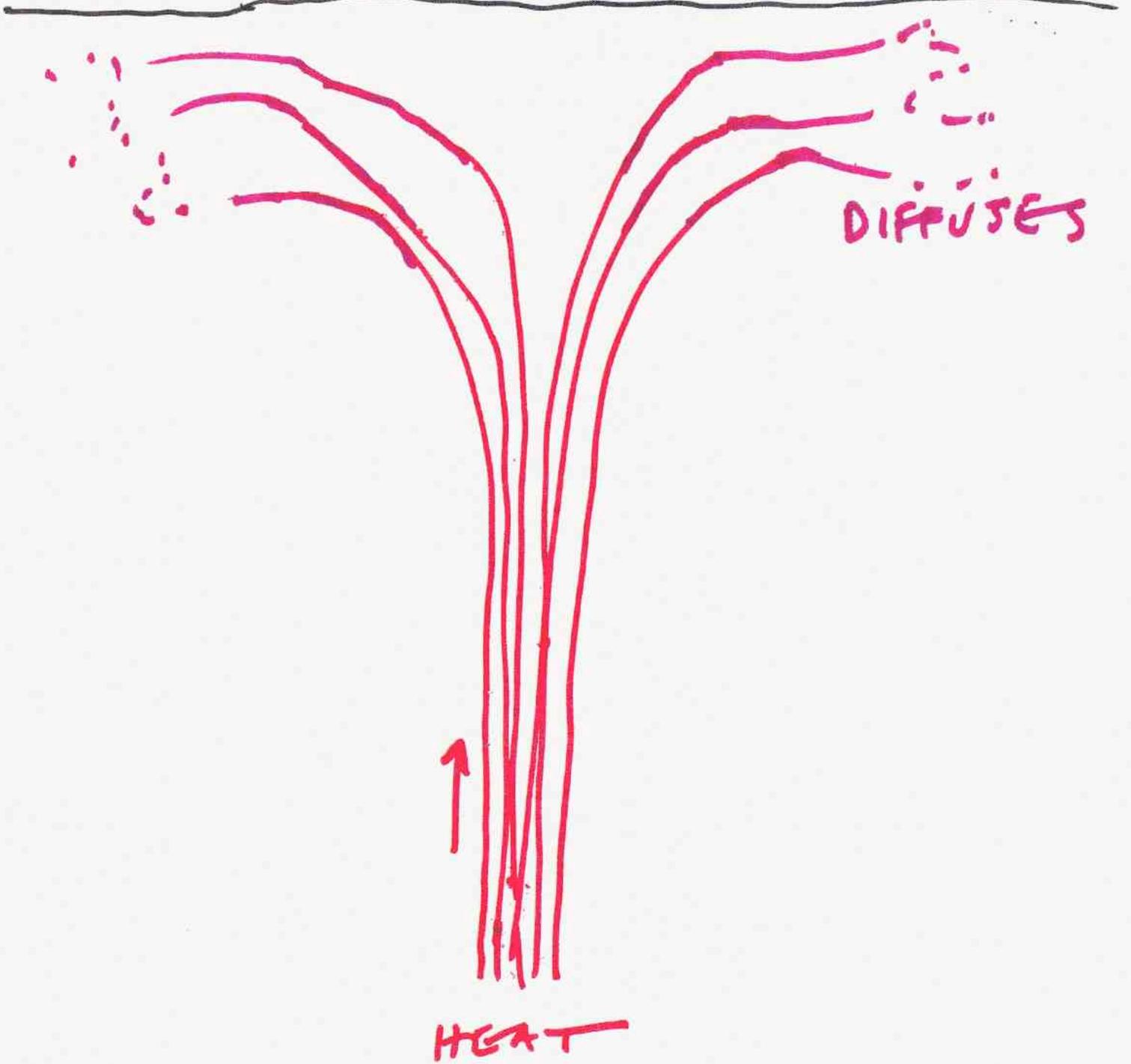

PLUME CONVECTION

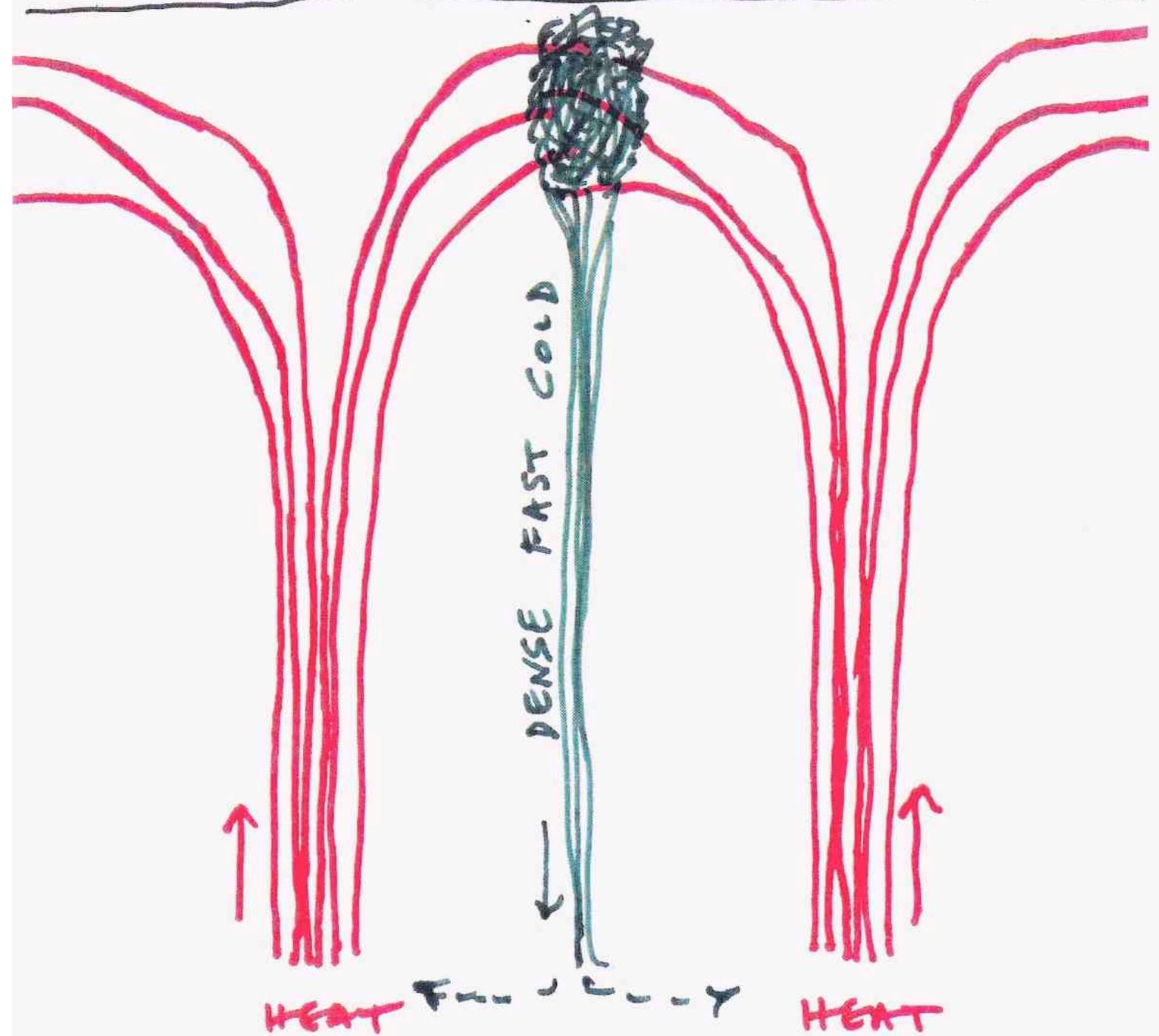

CONVECTION IN ATMOSPHERE AND ENVELOPE OF DWARF LATE TYPE STAR

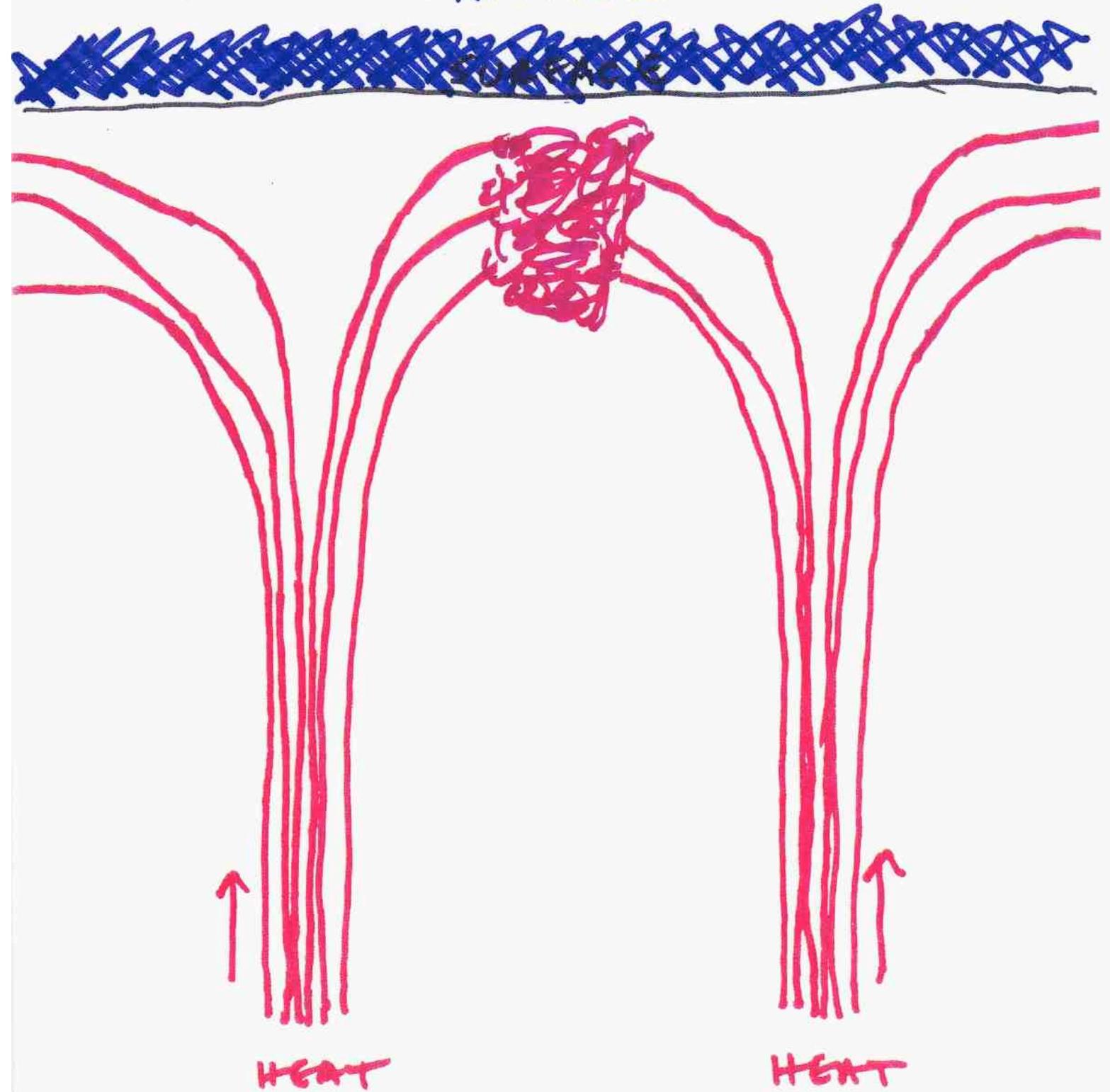

INFALL

SURFACE

HEAT   HEAT

INFALL STOPS RADIATIVE COOLING AND DOWNWARD FLOW BUT NOT UPWARD FLOW

# UNIFORMITARIANISM IN GEOLOGY

**100 YEARS AGO GEOLOGISTS WERE UNIFORMITARIANS. THERE WERE SLOW UPLIFTS, SLOW EROSION, SLOW LAYERING, AND THEN THE DEPOSITS WERE UPLIFTED AND THE CYCLE WAS REPEATED**

**ALL IMPORTANT CONSTRUCTION TOOK A LONG TIME AT MM/YEAR**

**IN 1915 WEGENER PROPOSED CONTINENTAL DRIFT THAT REQUIRED MOTIONS OF CM/YEAR AND UNKNOWN PHYSICS**

**THE UNIFORMITARIANS RIDICULED THE THEORY, SUPPRESSED RESEARCH ON IT, AND PREVENTED PROGRESS FOR MORE THAN A GENERATION**

**THE UNIFORMITARIANS ALSO SUPPRESSED CATASTROPHIC EXPLANATIONS FOR EVENTS, EVEN VIOLATING COMMON SENSE AND EVERYDAY EXPERIENCE**

**UNIFORMITARIANISM AND CONTINENTAL DRIFT WERE CONSISTANT AT ONE TIME WHEN ALL THE CONTINENTS MERGED INTO ONE, PANGEA, AND THERE WAS ONE OCEAN. THE MOUNTAINS WORE DOWN. THERE WAS NOT MUCH WEATHER AND NOT MUCH WEATHERING. EVERYTHING SLOWED DOWN.**

**FORTUNATELY, THE CORE IS RADIOACTIVELY HEATED AND CONVECTION EVENTUALLY BROKE UP PANGEA OR WE WOULD NOT BE HERE TODAY**

# UNIFORMITARIANISM IN BIOLOGY

**BIOLOGISTS USED TO THINK THAT EVOLUTION OCCURRED AT THE MUTATION RATE, RANDOMLY, SLOWLY, AND UNIFORMLY.**

**BUT, ACTUALLY, 90% OF EVOLUTION TAKES PLACE IN 10% OF THE TIME. "PUNCTUATED EVOLUTION".**

**EVOLUTION IS DRIVEN BY CATASTROPHES THAT PRODUCE LIFE AND DEATH SITUATIONS. SPECIES MUST EVOLVE OR DIE OUT WHEN THEIR ECOLOGICAL NICHE CHANGES OR DISAPPEARS.**

**NATURE DOES NOT BELIEVE IN SLOW EVOLUTION OF ANY SYSTEM**

**LONG BORING INTERVALS ARE ALWAYS PUNCTUATED BY EXCITING CATASTROPHES**

**THE CATASTROPHES ARE BUILT INTO SLOWLY EVOLVING SYSTEMS OR COME FROM OUTSIDE**

# UNIFORMITARIANISM IN ASTRONOMY

**CONVECTION IS UNIFORMITARIANISM AT WORK.  IT JUST GOES ON AND ON FOR BILLIONS OF YEARS FROM PROTOSTAR TO THE MAIN SEQUENCE AND UPWARD. BORING.**

**UNLESS THE MAGNETIC FIELD IN THE INTERIOR OF THE STAR HAS A SECRET LIFE, AT INTERVALS LONGER THAN PEOPLE HAVE BEEN OBSERVING.**

# FORMATION OF PLANETS

**THERE ARE SIMPLE, LOGICAL RULES THAT ARE THE SAME FOR EVERY UNARY LATE-TYPE POP I PROTOSTAR IN THE UNIVERSE**

**PLANETS ARE SELF-GRAVITATING BODIES > $10^{26}$ G FORMED FROM PROTOSTELLAR DISKS AFTER PROTOSTARS FORM**

**ALL UNARY DWARF LATE-TYPE STARS ARE FORMED FROM DISKS AND HAVE PLANETARY SYSTEMS LIKE OURS UNLESS AN OB STAR, OR A SUPERNOVA, OR A COLLISION HAS DISRUPTED THE DISK**

**BINARIES PROBABLY HAVE PLANETS AS WELL BUT I DO NOT DISCUSS THAT HERE.**

# QUESTIONS

**HOW ARE BINARIES MADE?**

**WHAT IS THE PURPOSE OF PLANETS?**

**WHY DOES A PROTOSTAR STOP GROWING?**

**WHAT HAPPENS TO THE DIPOLE MAGNETIC FIELD THAT COUPLES THE PROTOSTAR TO THE DISK?**

**WHAT HAPPENS TO INFALLING GAS ONCE THE MAGNETIC COUPLING IS BROKEN?**

# WASTE MANAGEMENT

**EVERY UNARY POP I PROTOSTAR HAS THE SAME WASTE MANAGEMENT PROBLEM**

## DISK WASTE PROPERTIES:

**RADIOACTIVE, SELF-HEATING**

**MAGNETIC, FE ~ SI**

**CARBONACEOUS**

**DAMAGE BY RADIOACTIVITY, RADIATION, AND PARTICLES**

**FREE RADICALS, BROKEN MOLECULES, POSITIVE AND NEGATIVE CHARGES**

**CARCINOGENIC**

**GENERALLY STICKY**

**FAR FROM STAR, ICY**

## DISPOSAL METHOD: DISPERSAL

**NEARBY OB STARS, OR SUPERNOVAS, OR COLLISIONS BLOW EVERYTHING AWAY.**
**IT IS FREE ENERGY.**

**DISK IS WIPED BEFORE STAR IS COMPLETED OR BEFORE PLANETS ARE COMPLETED**

**"URANUS" AND "NEPTUNE" ARE MOST LIKELY TO BE LOST BECAUSE THEY TAKE SO LONG TO MAKE**

## DISPOSAL METHOD: COMPACTION

**FIRST STEPS ARE TO PREVENT INFALL OF NEW WASTE AND TO GET WASTE IN DISK TO AGGLOMERATE SO IT IS EASIER TO HANDLE**

**PULLS THE DIPOLE MAGNETIC FIELD OUT OF THE DISK TO BREAK THE COUPLING**

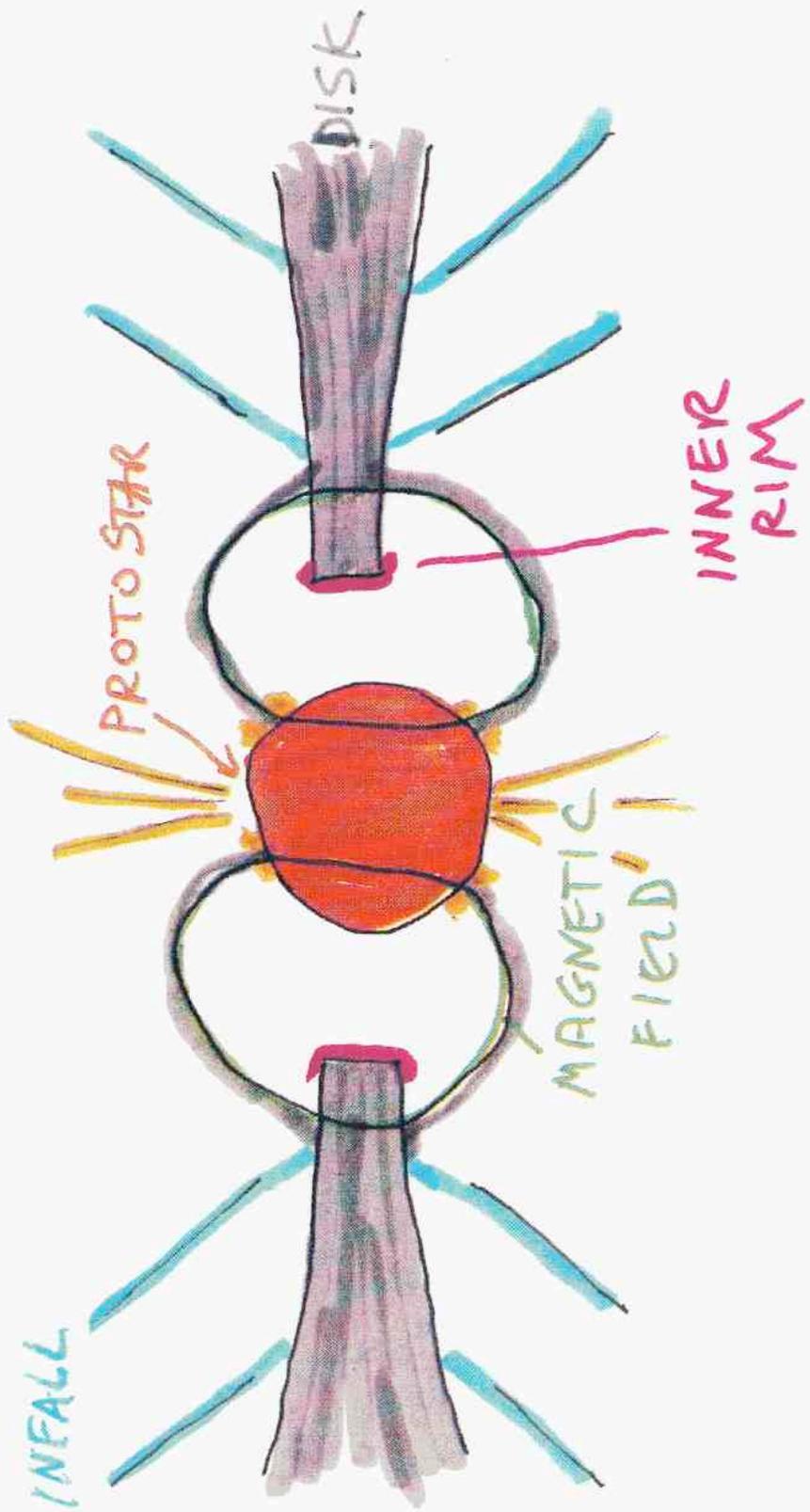

**THE PROTOSTAR ROTATES ONCE PER DAY**

**DIFFERENTIAL ROTATION AS IN THE SUN**

**PROTOSTAR IS SLIGHTLY OBLATE**

RADIATION PRESSURE PUSHES DISK MATERIAL AWAY AND FORMS A WALL OF DENSE, WARM GAS AND DUST THAT ABSORBS 10 PER CENT OF THE RADIATION FROM THE STAR.

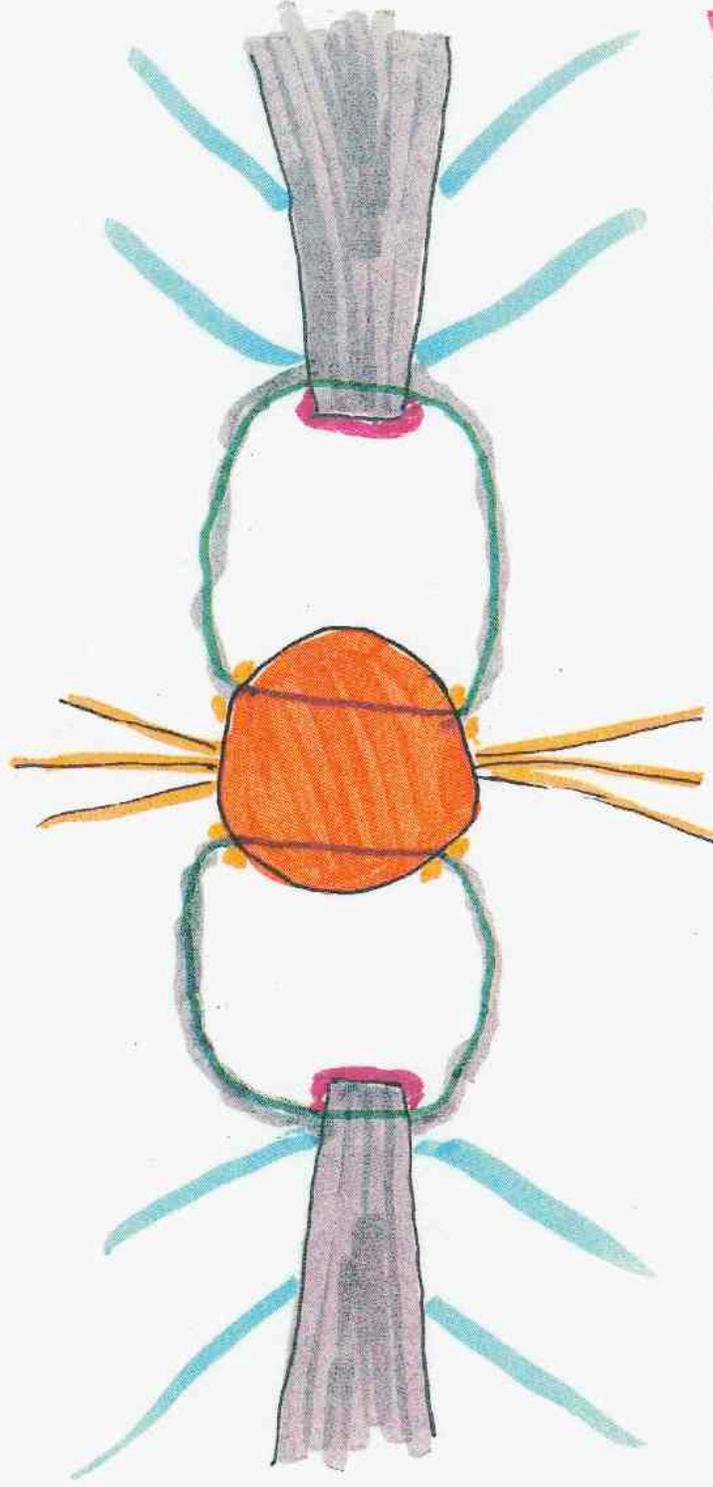

MAGNETIC FIELD **IS** STRETCHED OUTWARDS WITH THE DISK SINCE IT IS PINNED TO THE DISK.

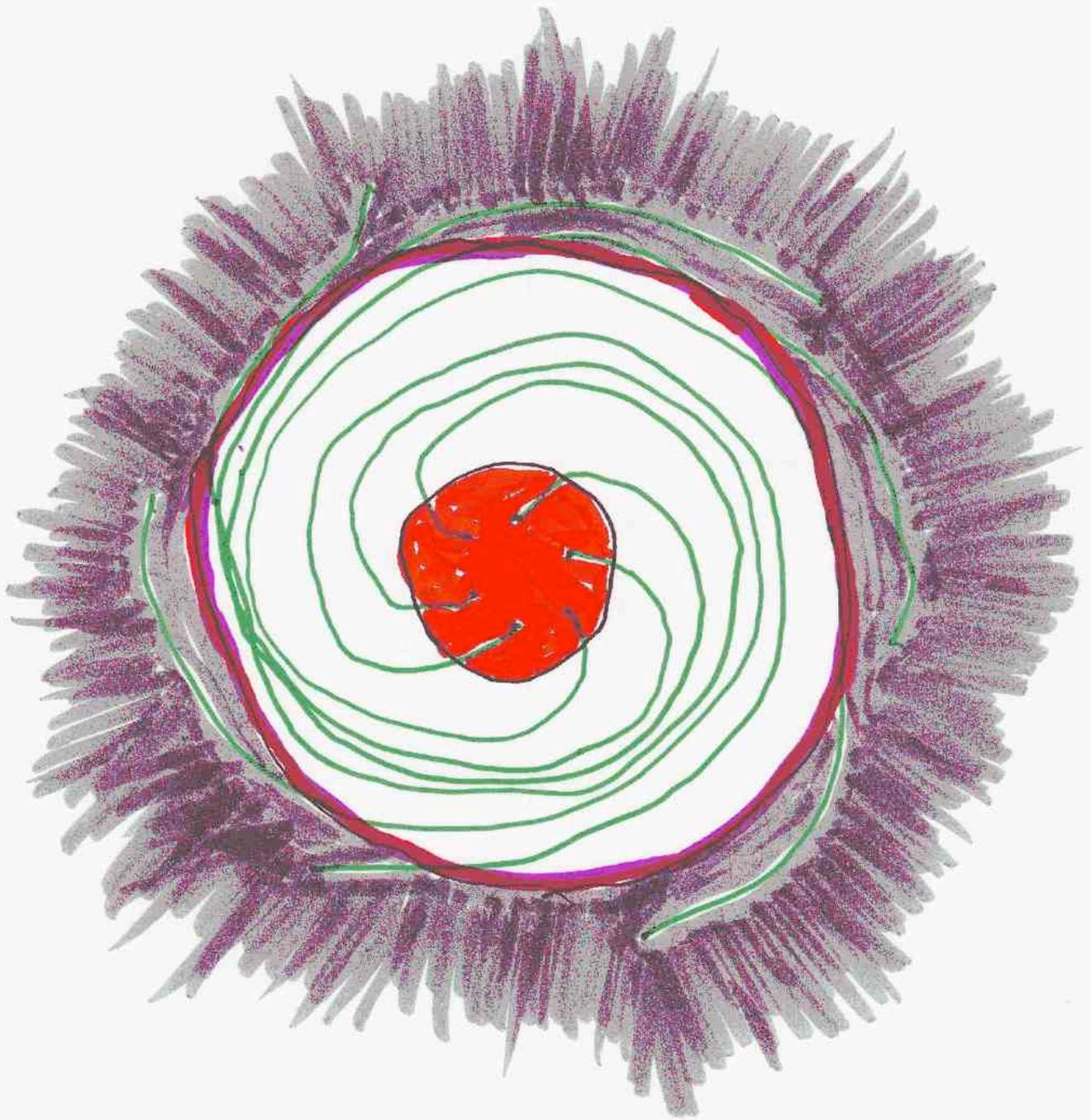

# WHAT HAPPENS TO THE DIPOLE MAGNETIC FIELD?

WHEN THE DIPOLE FIELD PULLS OUT OF THE DISK TO FORM THE SUPER-CME, IT FILLS THE SURFACE OF THE STAR FROM ABOUT +70 TO -70 DEG LAT

**SUPER-CME ILLUMINATES BOTH SIDES OF THE DISK AS WELL AS THE INNER WALL**

RADIATION AND PARTICLES BLOW AWAY ALL THE INFALLING GAS OUT TO ~ SNOW LINE AND PUSH INWARD ON BOTH SURFACES OF THE DISK SO THAT DUST AGGLOMERATES. VAPORIZES AT SURFACE

**OPTICAL DEPTH OF POPCORN IS MUCH LESS THAN THAT OF DUST FROM WHICH IT IS MADE**

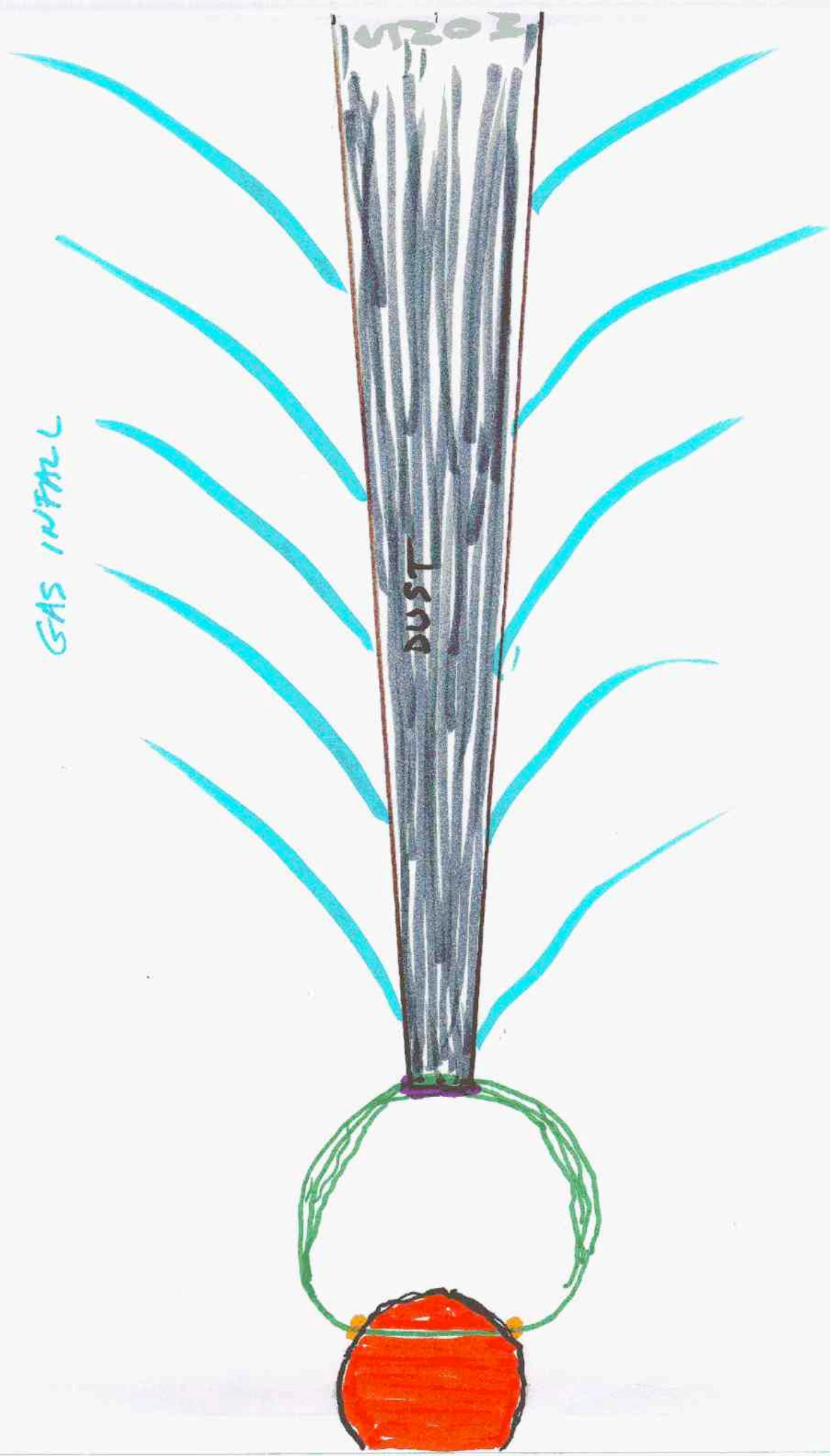

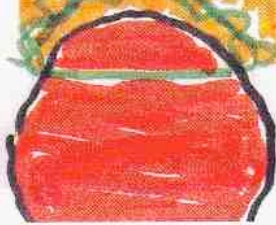

FILLED WITH RADIATION + WIND

RADIATION + WIND

SUPER CME

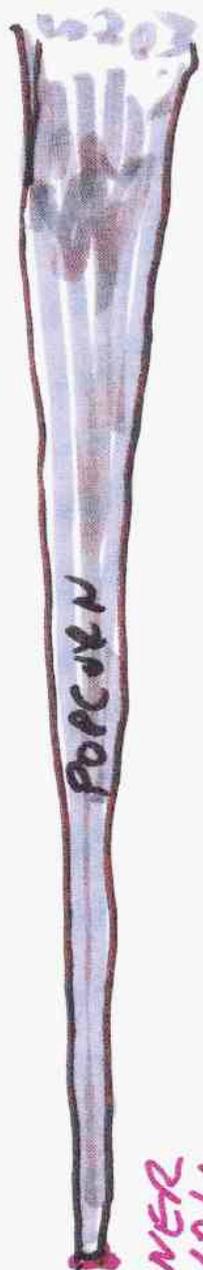
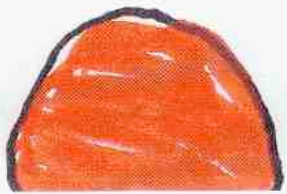

**PLANETS FORM SEQUENTIALLY OUTWARD**

**RADIATION AND WIND FROM STAR HEAT POPCORN AND PUSH IT OUTWARD**

**ORBITS MOVING OUTWARD CROSS STATIONARY ORBITS AND ORBITS MOVING INWARD AT AT SHALLOW ANGLE SO SLOW COLLISIONS**

**WALL OF AGGLOMERATED POPCORN (= POPCORN BALLS) FORMS THAT ABSORBS 3-5% OF THE RADIATION FROM THE STAR**

**NEAR STAR VOLATILES ARE COOKED OUT**

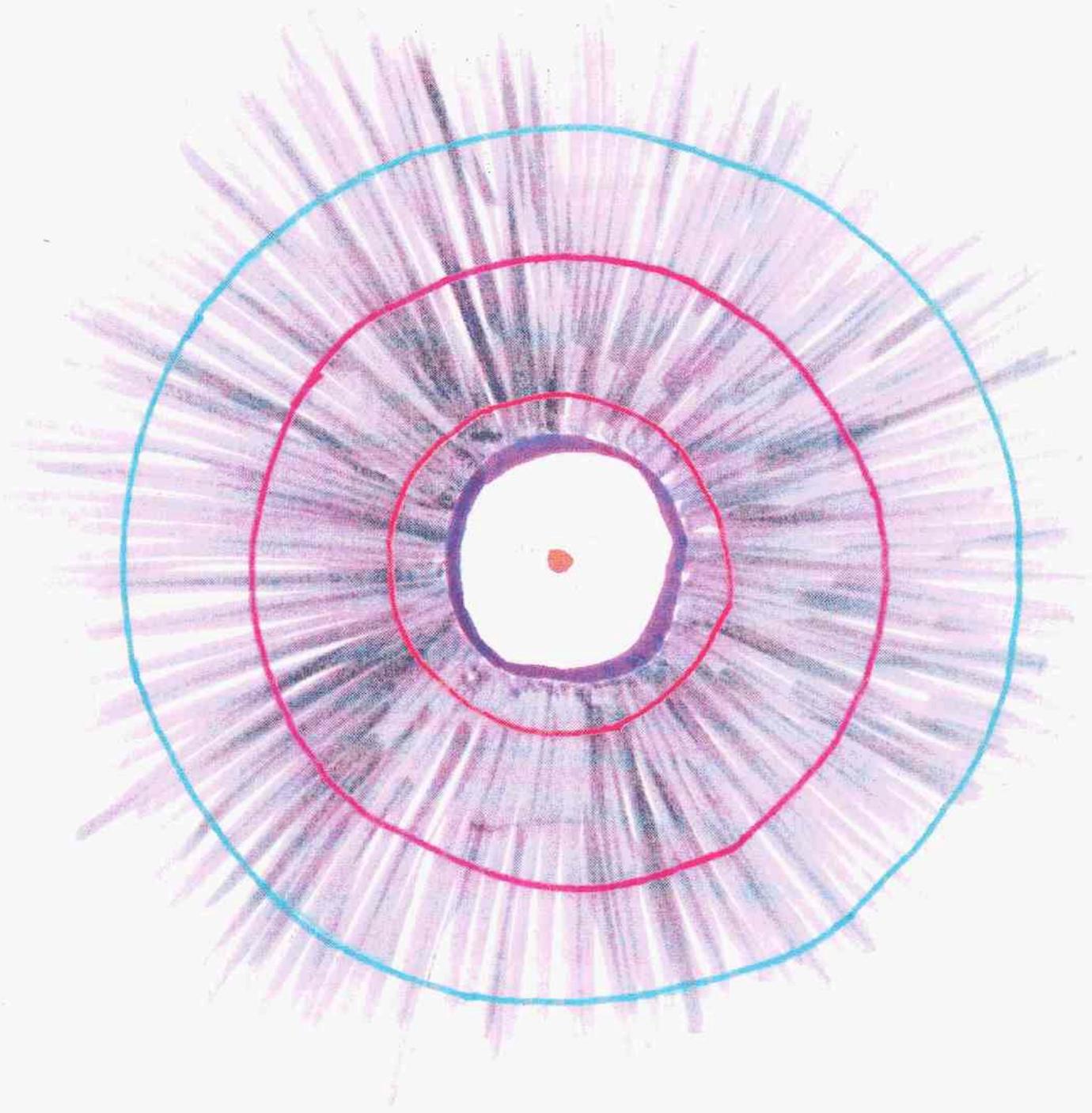

**POPCORN AGGLOMERATES AND CONSOLIDATES UNTIL PLANETESIMALS FORM AND THE WALL BECOMES OPTICALLY THIN**

**A NEW WALL FORMS FURTHER OUT WHERE THE DISK IS INTACT AND THE PROCESS REPEATS**

**AS THE RADIUS INCREASES THERE IS LESS "COOKING" OF THE POPCORN**

# POPCORN

STICKY
CHRUSHABLE
INSULATING
RADIOACTIVELY SELF-HEATING
WARMED AND ANNEALED BY STAR
NOT LIKE PARTICLES IN
MANY BODY CALCULATIONS

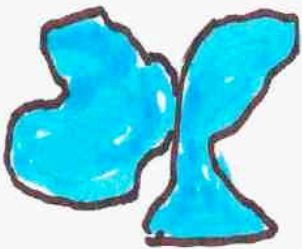

LOW VELOCITY
COLLISION
STICKS

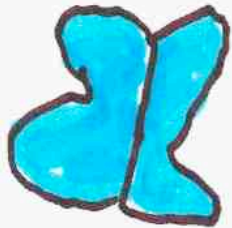

MODERATE
VELOCITY COLLISION
CRUSHES

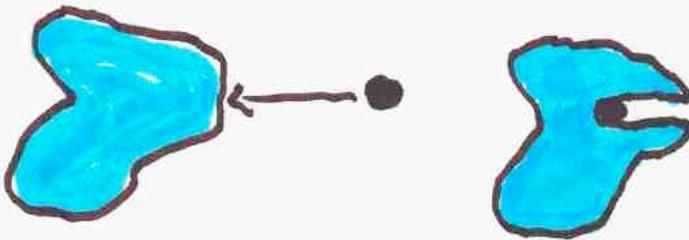

COLLISION WITH SMALL DENSE
FAST PARTICLES PENETRATE

WHEN BODY GROW LARGER HEAT FROM RADIOACTIVITY IS TRAPPED IN THE CENTER AND DIFFERENTIATION BEGINS.

WHEN GROWS TO SELF-GRAVITATING PLANETESIMAL CRUSHES UNDER OWN WEIGHT. DIFFERENTIATION INCREASES. MELTING BEGINS. DIFFERENTIATION IS MAINTAINED WHEN PLANETESIMAL MERGE.

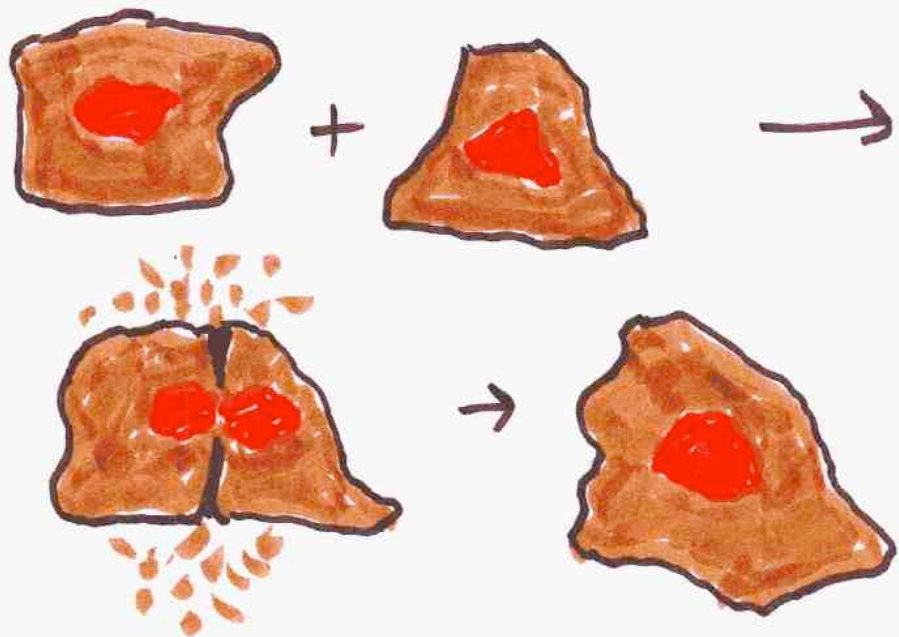

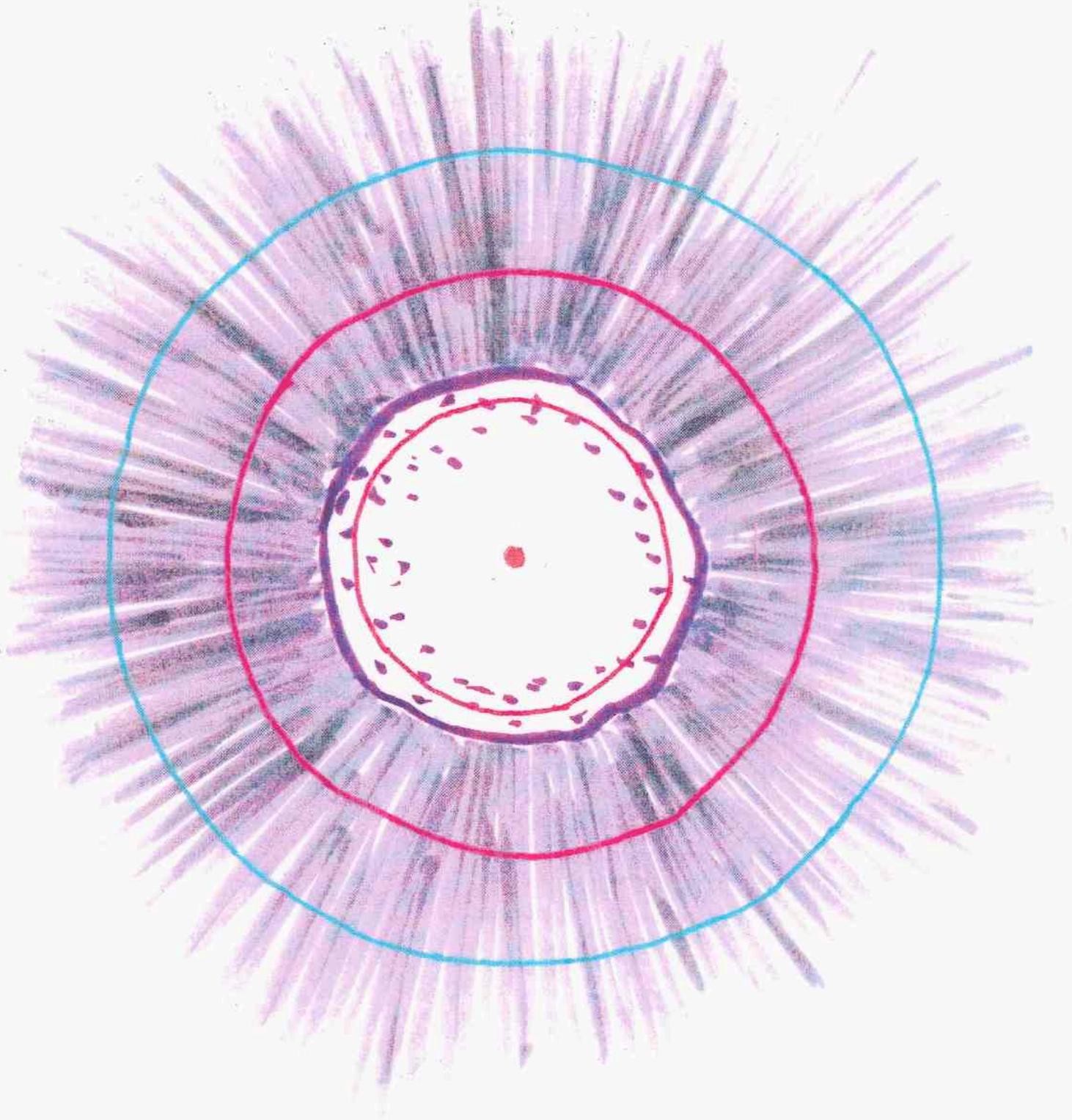

**PLANETESIMALS FORMED IN THE WALL ARE IN NEARBY ORBITS**

**THE DISPERSION OF THE DOTS IN THE CARTOONS IS GREATLY EXAGGERATED**

**THEY RAPIDLY GRAVITATIONALLY COLLAPSE INTO A PLANET**

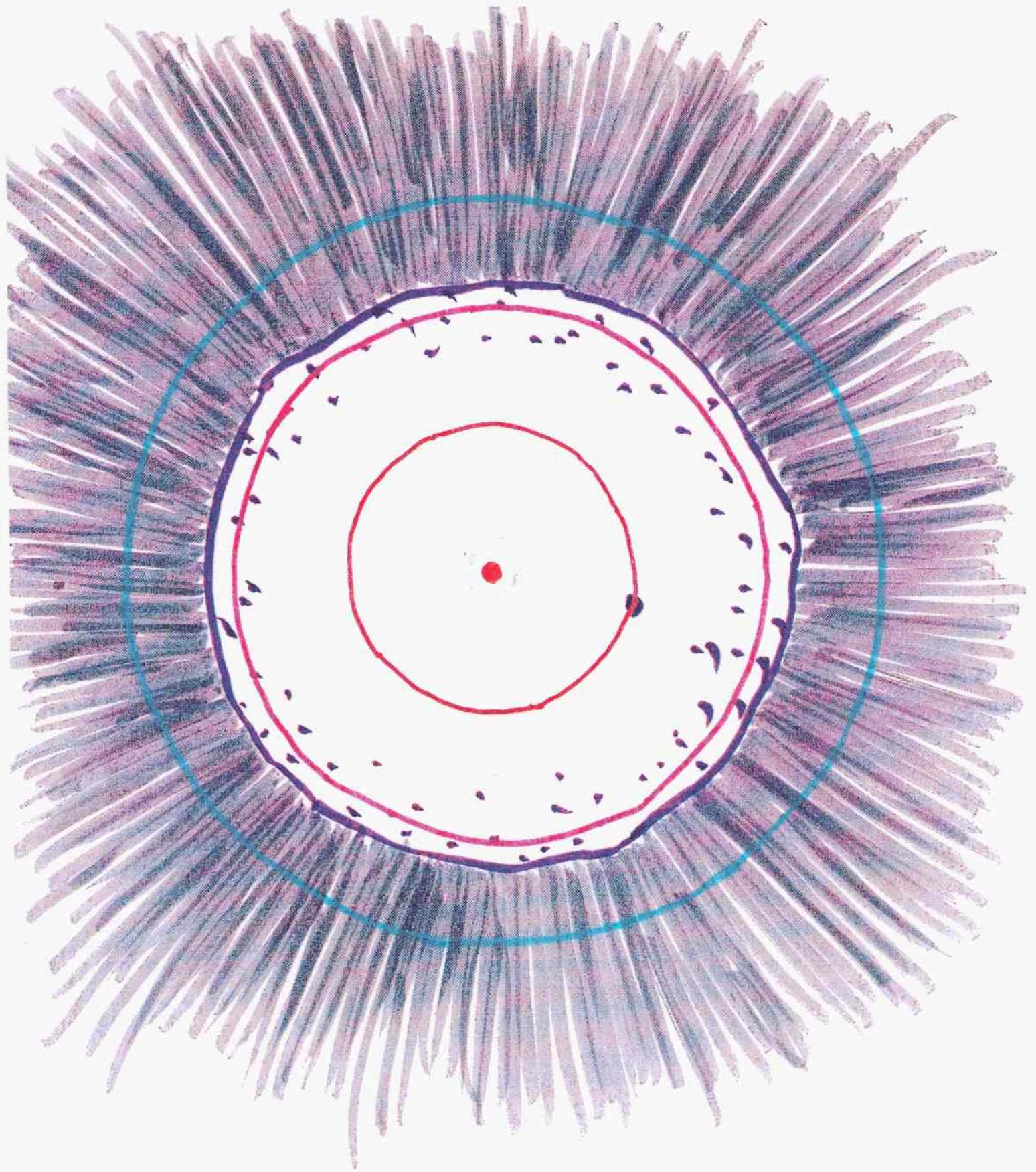

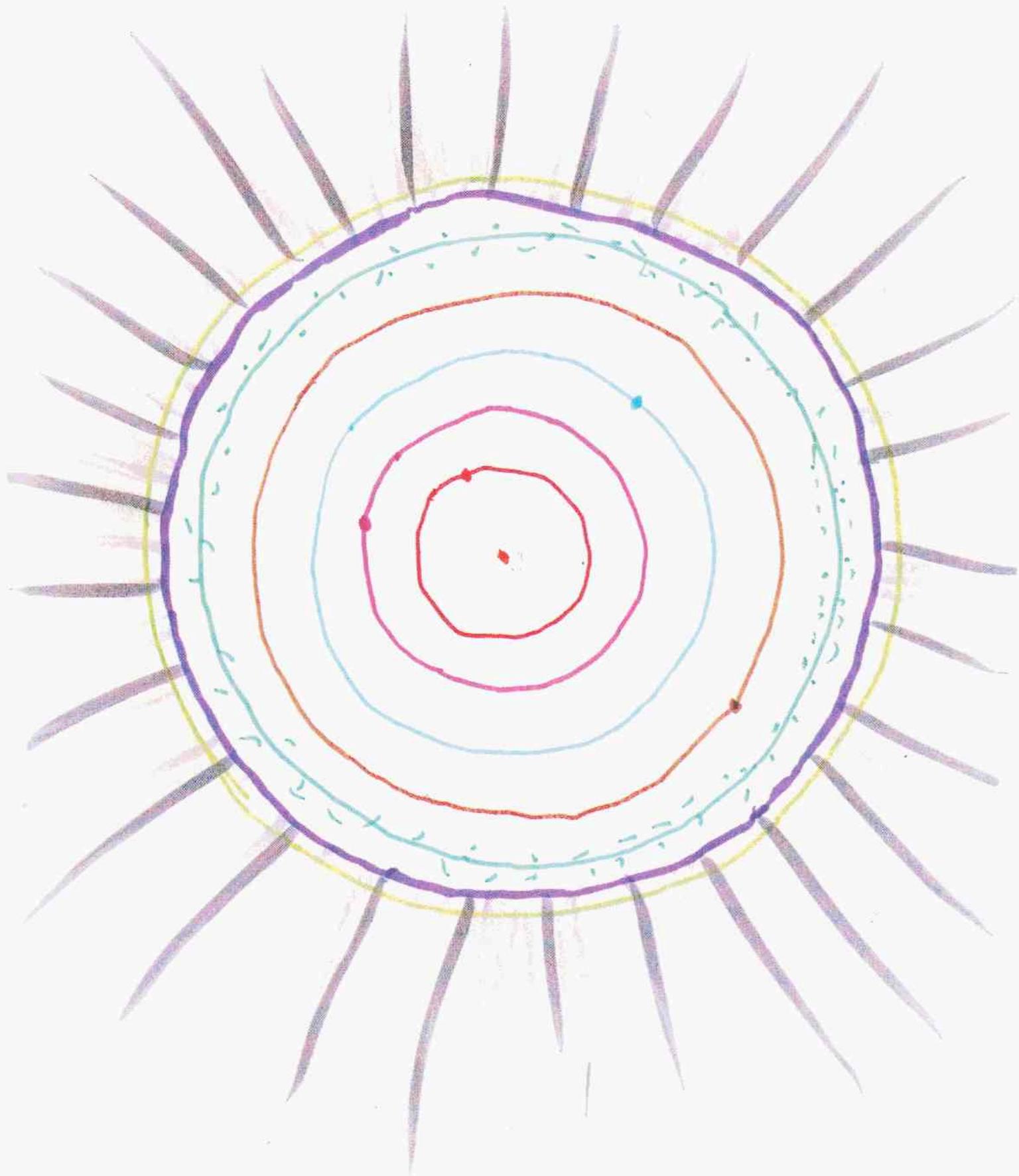

**NOT ENOUGH MASS IN ASTEROID WALL TO FORM A PLANET. LEFT WITH PLANETESIMALS**

**SNOW BELT**

**ICE IS ADDED TO THE POPCORN AND DUST IN JUPITER'S WALL. A LARGE CORE FORMS THAT IS ABLE TO PULL IN GAS**

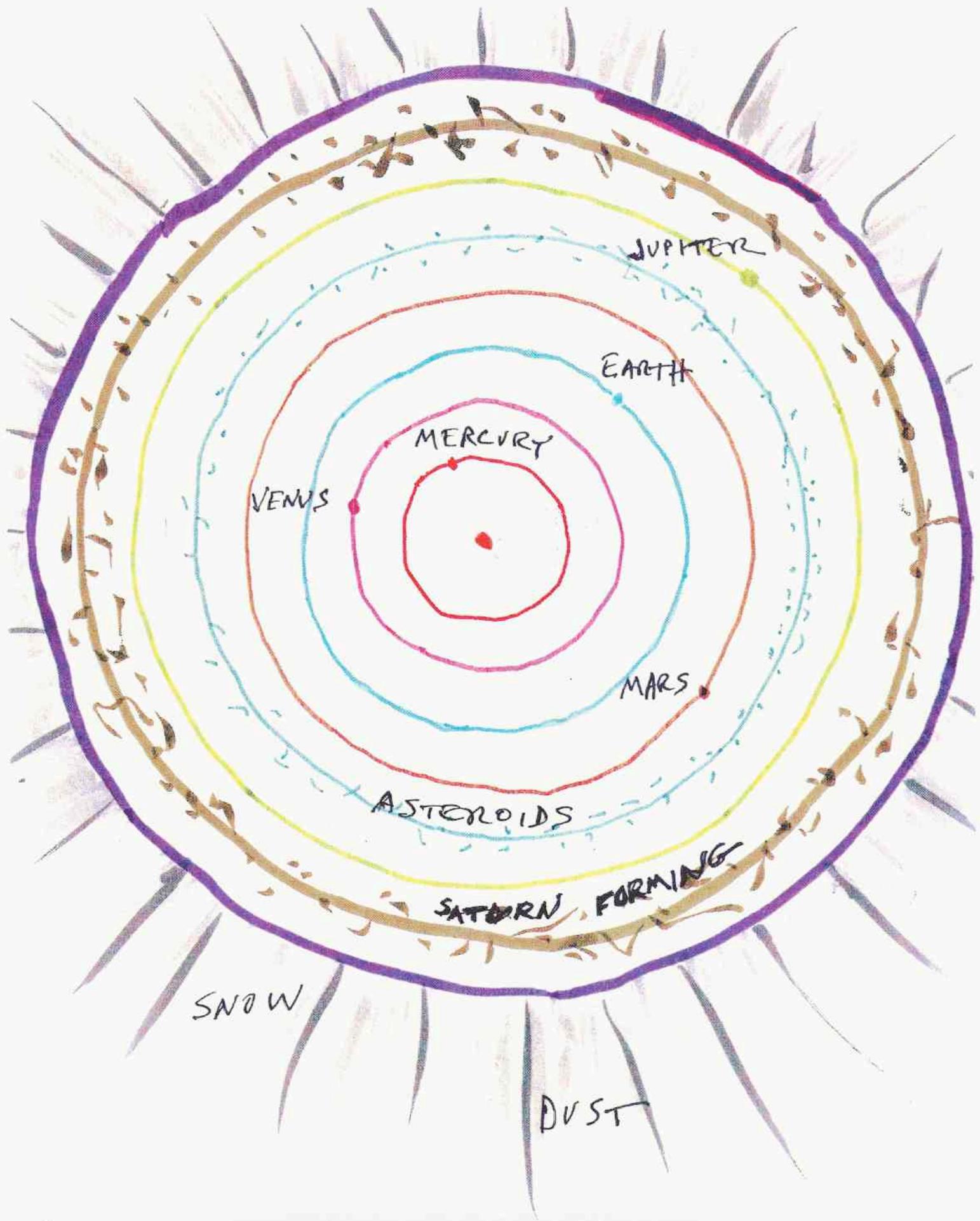

## UNIFORMITARIANISM WORKS FOR THE OUTER PLANETS

## SINCE RADIATION FROM THE STAR IS SO WEAK AT THOSE DISTANCES, THEY ARE ACCRETED SLOWLY BY UNASSISTED GRAVITY

```
REMINDER: CFA Colloquium 4 May 2006

SPEAKER:   David Wilner (CfA)

TITLE:     Protoplanetary Disks:
           Analogs of the Early Solar System
```

```
ABSTRACT: [edited]
```

The nearly circular, coplanar orbits of the planets in our Solar System have led astronomers for centuries to consider an origin within a rotating disk.

Advances in observational resolution and sensitivity, especially at millimeter wavelengths, now provide direct evidence for disks of gas and dust surrounding nearby pre-main-sequence stars, many of which have physical properties appropriate to the formation of Solar Systems like our own.

I will review briefly how we know what we know about these "protoplanetary" disks, and I will discuss how new observations are starting to capture key aspects of the planet formation process, including

(1) dust evolution towards "pebbles", and

(2) inner disk holes opened by disk-protoplanet interactions.

**DAVE WILNER SHOWED IMAGES OF THE 20 CLOSEST "PROTOPLANETARY" DISKS THAT ARE VISIBLE THROUGH SURROUNDING GAS AND DUST**

**THE DISKS HAD UNRESOLVED CENTRAL HOLES WITH DIAMETERS > 20 AU.  SMALLER HOLES ARE NOT VISIBLE**

**IN FACT, THESE ARE "POST-PLANETARY" DISKS.  PLANETS HAVE ALREADY FORMED RAPIDLY OUT THROUGH "JUPITER"**

SPEAKER:   Renu Malhotra (University of Arizona)

TITLE:     Bombardment History of the Terrestrial Planets

ABSTRACT:  [edited]

Analysis of the impact crater record of the terrestrial planets and the Moon implicates two populations of impactors that are distinguishable by their size distributions.

The old population, responsible for an intense period of bombardment that ended 3.8 Gy ago, is virtually identical in size distribution to the present main belt asteroids;

the second population, responsible for craters younger than ~3.8 Gy, matches closely the size distribution of the near earth asteroids.

An inner Solar System impact cataclysm occurred ~3.9 Gy ago,

Many asteroids were ejected from the main belt on a short timescale, ~$10^7$ y, possibly due to a major dynamical instability owed to the orbital migration of the giant planets in early Solar System history.

# FORMATION OF OUTER PLANETS

**RENU MALHOTRA SHOWED THAT URANUS AND NEPTUNE TOOK ~600 MILLION YEARS TO FORM AND MIGRATE INTO PLACE**

**THEIR FORMATION AND MIGRATION PRODUCED RESONANCES IN THE ASTEROID BELT THAT EITHER DROVE ASTEROIDS INWARD TO CRATER THE INNER PLANETS, OR EJECTED THEM FROM THE SOLAR SYSTEM**

# FORMATION OF CRYPTOPLANETS

```
DATE: MONDAY, November 20
TIME: 11:30
ROOM: Pratt
SPEAKER:  Frederic Pont (Geneva Observatory)
```

"Transiting Extra-solar Planets:
 Towards Exo-planetology"

Abstract [edited]

It is now "Year 11" for extrasolar planet studies, and the frantic pace of progress shows no sign of slowing down.

Recent landmarks since the last Cool Star meeting include the detection of planets of a few Earth masses by radial velocity and microlensing surveys, and the characterization of several transiting hot gas giants.

Transiting planets have served as the Rosetta Stone to connect the abstract language of exoplanet orbits to the detailed knowledge of solar system planets, providing an empirical foundation to the nascent field of "exo-planetology".

Precise masses and radii are known for more than a dozen transiting gas giants, some discovered by wide-field photometric monitoring, others by radial velocity surveys.

For the brightest of these, infrared emission from the planet itself could be detected at several wavelengths with the Spitzer Space Telescope.

The Corot satellite is expected to discover dozens of transiting planets, possibly including ice giants and rock giants.

**PONT'S TRANSITING PLANET TABLE**

| NAME | Mplanet | Rplanet | Period | a [AU] |
|---|---|---|---|---|
| OGLE-TR-10 | 0.61 | 1.22 | 3.101278 | 0.04162 |
| OGLE-TR-56 | 1.29 | 1.30 | 1.211909 | 0.0225 |
| OGLE-TR-111 | 0.52 | 1.01 | 4.0144479 | 0.0467 |
| OGLE-TR-113 | 1.35 | 1.09 | 1.4324757 | 0.0229 |
| OGLE-TR-132 | 1.14 | 1.18 | 1.689868 | 0.0299 |
| HD189733 | 1.15 | 1.156 | 2.2185733 | 0.031 |
| HD149026 | 0.330 | 0.726 | 2.87598 | 0.042 |
| TrES-1 | 0.76 | 1.081 | 3.0300737 | 0.0393 |
| TrES-2 | 1.28 | 1.24 | 2.47063 | 0.0367 |
| HD209458 | 0.657 | 1.320 | 3.52474859 | 0.047 |
| XO-1 | 0.90 | 1.184 | 3.941534 | 0.0488 |
| HAT-P-1 | 0.53 | 1.36 | 4.46529 | 0.0551 |
| WASP-1 | 0.867 | 1.443 | 2.519961 | 0.0382 |
| WASP-2 | 0.88 | 1.038 | 2.152226 | 0.0307 |

| NAME | Mstar | Rstar | Teff | [Fe/H] |
|---|---|---|---|---|
| OGLE-TR-10 | 1.10 | 1.14 | 6075 | 0.28 |
| OGLE-TR-56 | 1.17 | 1.32 | 6119 | 0.19 |
| OGLE-TR-111 | 0.81 | 0.831 | 5044 | 0.19 |
| OGLE-TR-113 | 0.78 | 0.77 | 4804 | 0.15 |
| OGLE-TR-132 | 1.26 | 1.34 | 6210 | 0.37 |
| HD189733 | 0.82 | 0.753 | 5050 | −0.03 |
| HD149026 | 1.3 | 1.45 | 6147 | 0.36 |
| TrES-1 | 0.89 | 0.811 | 5250 | −0.02 |
| TrES-2 | 1.08 | 1.00 | 5960 | 0 |
| HD209458 | 1.101 | 1.125 | 6117 | 0.02 |
| XO-1 | 1.0 | 0.928 | 5750 | 0.015 |
| HAT-P-1 | 1.12 | 1.15 | 5975 | 0.13 |
| WASP-1 | 1.15 | 1.453 | 6200 | ? |
| WASP-2 | 0.79 | 0.813 | 5200 | ? |

**PONT SAID, IN JEST, THAT EXOPLANETS SEEM TO BE DIFFERENT FROM OUR PLANETS AND THAT OUR PLANETS WOULD TURN OUT TO BE WEIRD**

**THAT IS WHY I DECIDED TO GIVE THIS TALK**

# SAO detects 'puffy' new planet

BY CAROLYN MARTIN
OPA Staff Writer

The discovery of a huge, bizarre, "puffy" new planet and the announcement of a new geological identification chart for planets garnered international headlines in September, after scientists from the Smithsonian Astrophysical Observatory briefed journalists at a news conference in the Castle Commons.

The newly discovered planet, called HAT-P-1, is the largest ever found inside or outside our solar system, and it has the lowest density of any planet ever measured. It orbits a distant star about 450 light-years away in the constellation Lacerta.

Robert Noyes, SAO research astrophysicist, and Gaspar Bakos, Hubble fellow at SAO, discovered the planet using a network of small, automated telescopes.

HAT-P-1 has a radius 1.38 times Jupiter's, but its mass is only half that of Jupiter, and it is less dense than water—astronomers affectionately describe it as "puffy," Noyes explained. "If you put this planet in a giant tub of water, it would float on the surface like a cork ball."

The new planet is named for the telescopes scientists used to discover it—the HAT network of six small, automated telescopes at Smithsonian facilities in Arizona and Hawaii. The telescopes search for planets by watching for stars that dim slightly when an orbiting planet crosses directly in front of the star as viewed

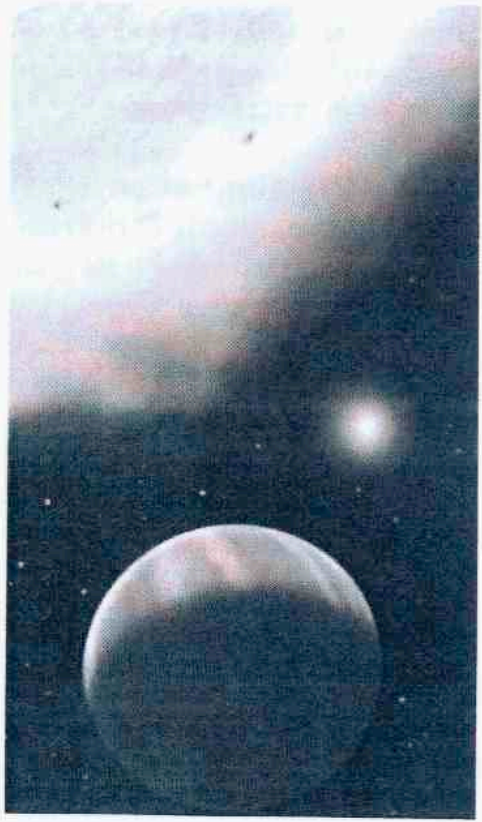

HAT-P-1, the newly discovered planet, orbits a star 450 light-years away. Although it is the largest planet ever measured, it would float if there was a bathtub large enough to hold it. (Image by David Aguilar)

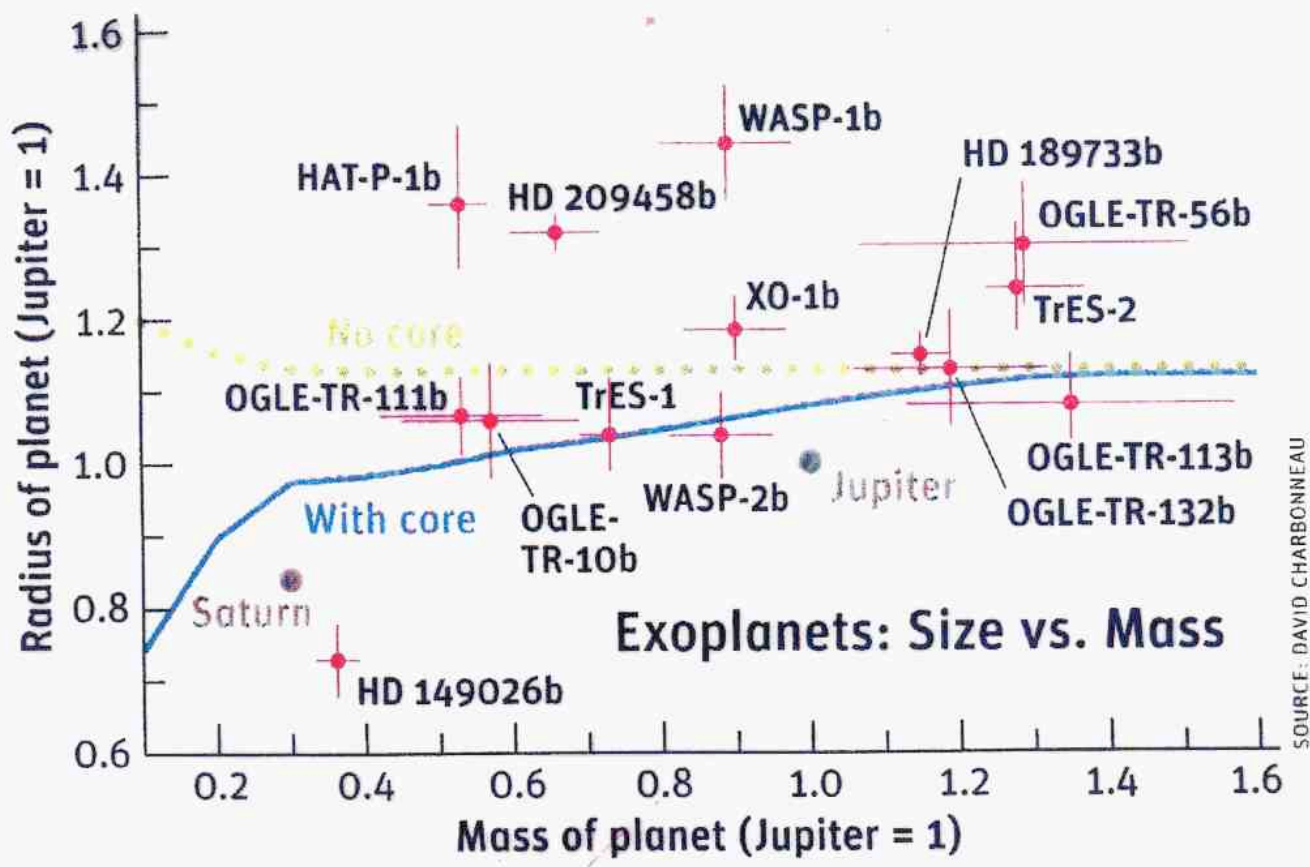

Exoplanets: Size vs. Mass

This graph plots the masses and radii of 14 known transiting exoplanets, along with Jupiter and Saturn. At least four planets, and possibly as many as five, have radii that are much larger than theory predicts. The lines represent theoretical models for close-in planets with a 20-Earth-mass core and with no core.

FROM DAVE CHARBONNEAU

"I think we're missing something fundamental about the interior structure or atmospheres of hot Jupiters."

--- Josh Winn

```
DATE: MONDAY, October 30
TIME: 12:00
ROOM: Pratt
SPEAKER: Jason Wright (UC Berkeley)
```

"Characterizing Exoplanetary Systems:
The Search for Solar System Analogs"

Abstract [edited]

For 10 years the California & Carnegie Planet Search
has been collecting precision radial velocities of
hundreds of Sun-like stars with ever-improving precision.

Today, this temporal baseline allows us to detect
exoplanets at orbital distances of ~ 5 AU including
many in systems already known to harbor an inner
exoplanet.

Analysis of incomplete orbits of longer-period
exoplanets allows us to peek beyond this 5 AU curtain,
and recent hardware and software upgrades bring our RV
precision to 1 m/s, allowing us to detect exoplanets
of only a few earth masses.

Together, these achievements have helped bring the
number of known multiple-planet systems up to 20
(and counting) and test theories of planet formation
and migration.

Although it appears that multiplicity among exoplanets
is common, the search for a true Solar System analog
continues.

# EXOPLANETS

**THE ONLY CONNECTION BETWEEN PLANETS AND EXOPLANETS IS THAT THEY BOTH ORBIT STARS**

**IN GENERAL:**

**EXOPLANETS ORBIT CLOSE TO THE STAR**
**EXOPLANETS HAVE MORE ECCENTRIC ORBITS**
**EXOPLANETS ARE MORE MASSIVE**
**EXOPLANETS ARE LESS DENSE**

**NOT A SINGLE EXOPLANET CAN BE MATCHED TO A PLANET**

**ANYONE WITH COMMON SENSE WOULD THINK THAT THEY ARE TWO SEPARATE CLASSES OF OBJECT**

**A NEW CLASS OF OBJECT REQUIRES AN EXPANATION AND IMAGINATION**

**I DID NOT BELIEVE THAT THE FIRST EXTRA SOLAR GIANT PLANET WAS A PLANET.  NEITHER DID DAVID BLACK.**

**I MADE UP A NAME FOR THE CLASS: CRYPTOPLANETS**

**WHEN OBSERVATIONS IMPROVE THERE WILL EVENTUALLY BE EXOPLANETS THAT ARE PLANETS, SO EXOPLANET IS NOT A VALID NAME FOR THE CLASS**

# CRYPTOPLANETS

**WHAT ARE THEY?**

**PLANETS ARE WASTE DUMPS FROM THE DISK**

**MAYBE THEY ARE WASTE DUMPS, BUT NOT FROM THE DISK**
- **WASTE INFALL?**
- **WASTE ANGULAR MOMENTUM?**
- **WASTE MAGNETIC FIELD?**

**SINCE CRYPTOPLANETS ARE CLOSE TO THE STAR, THEY ARE PROBABLY ALL THREE AND FORM BEFORE PLANETS**

**START WITH THE SAME PROTOSTAR MACHINERY AS BEFORE**

**STANDARD SCENARIO ASSUMES ISOLATED EVOLUTION BUT STARS ARE FORMED IN CLUSTERS**

**IF THERE IS A COLLISION WITH A DENSE CLOUD, OR WITH A STAR, OR IF AN OB STAR FORMS NEARBY, OR THERE IS A SUPERNOVA, THERE CAN BE A PULSE OF ENHANCED DENSITY**

**LET THE COMPLEX MACHINERY NOT WORK PERFECTLY SO THAT THERE IS EXCESS INFALL THAT THE DIPOLE MAGNETIC FIELD CANNOT CONTROL**

**INFALL GAS FLOWS OVER THE SURFACE OF THE STAR TO THE EQUATOR AND THEN OUTWARD**

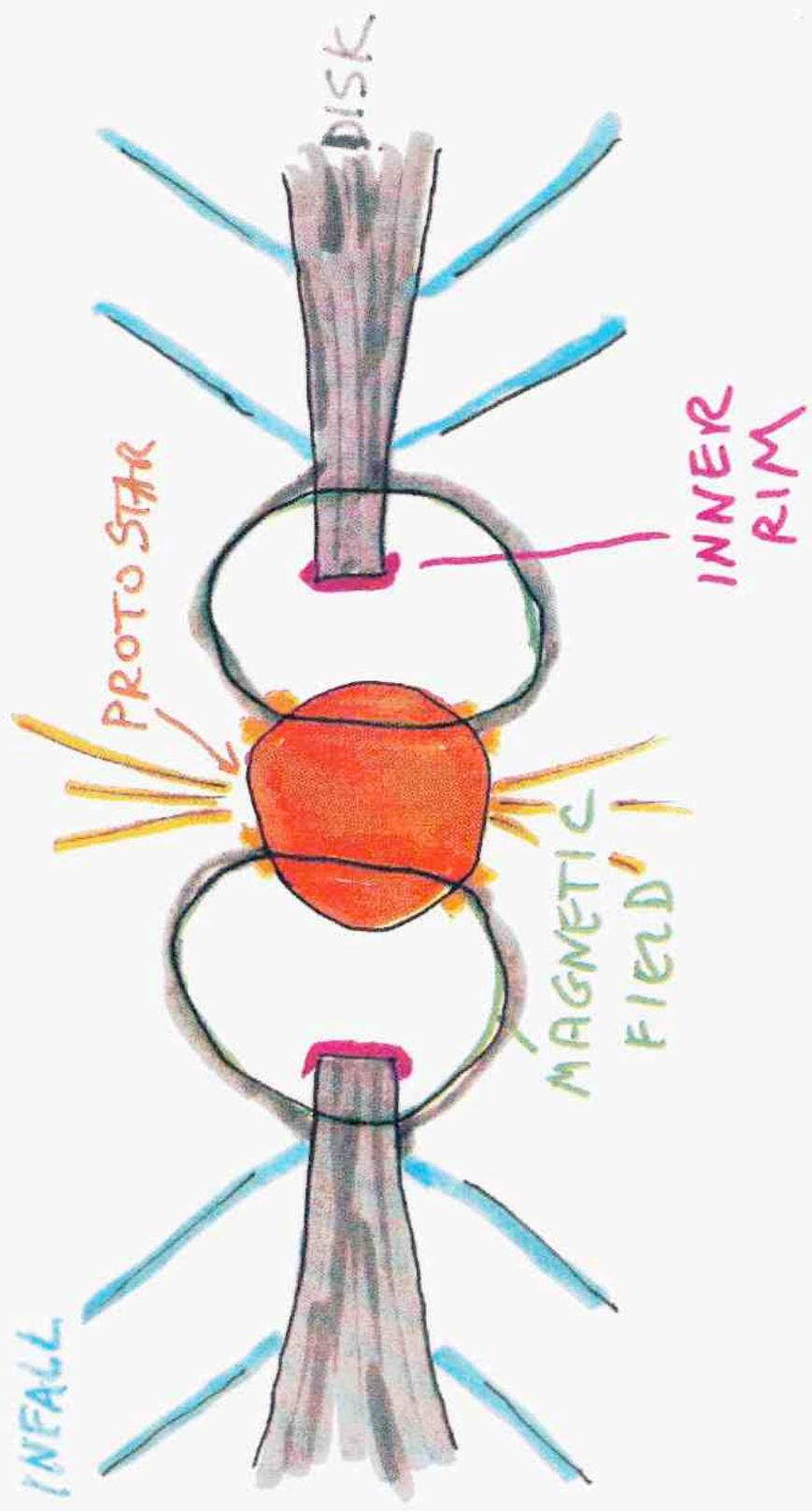

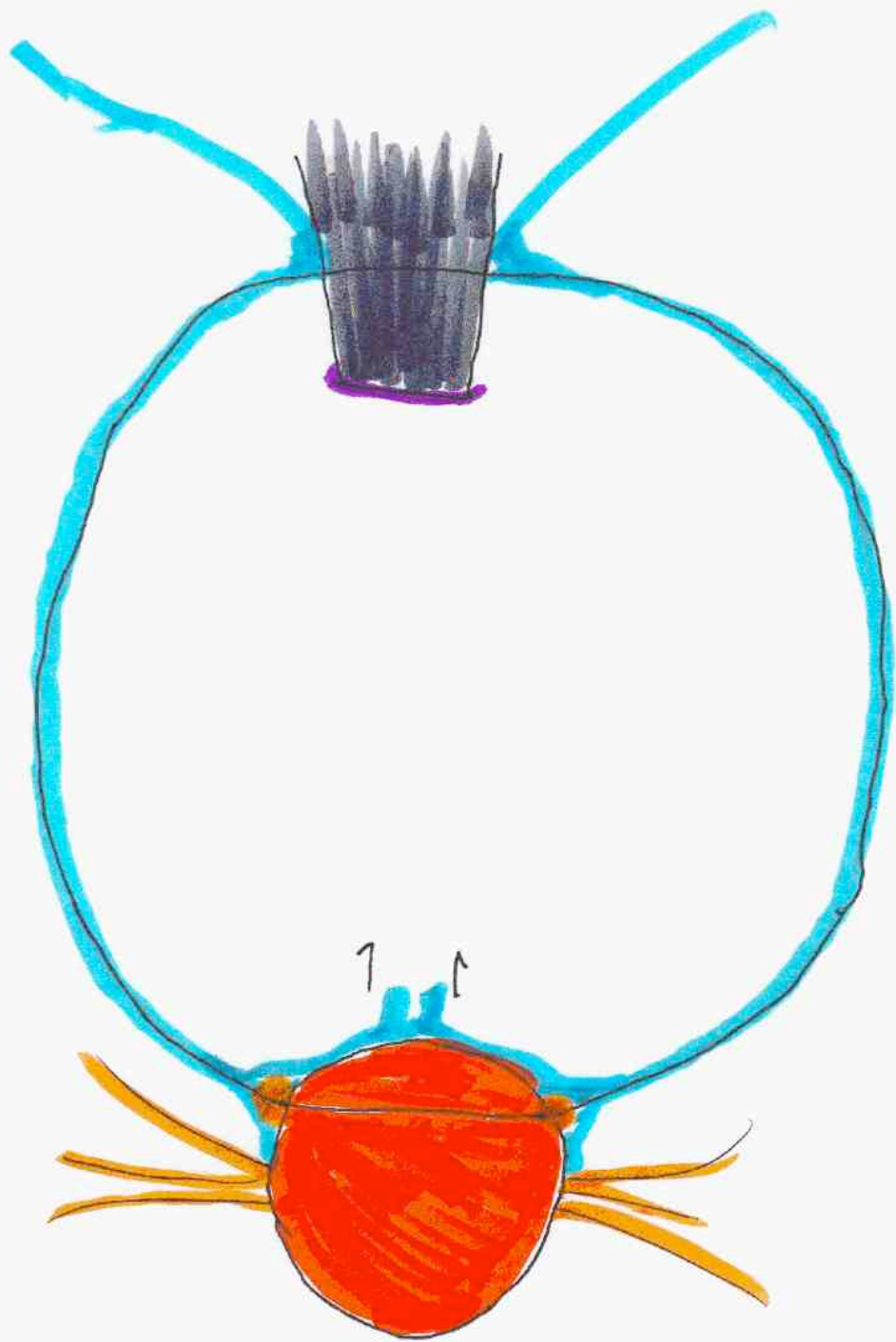

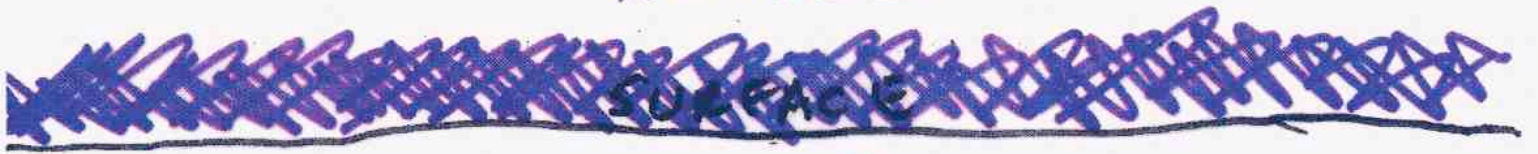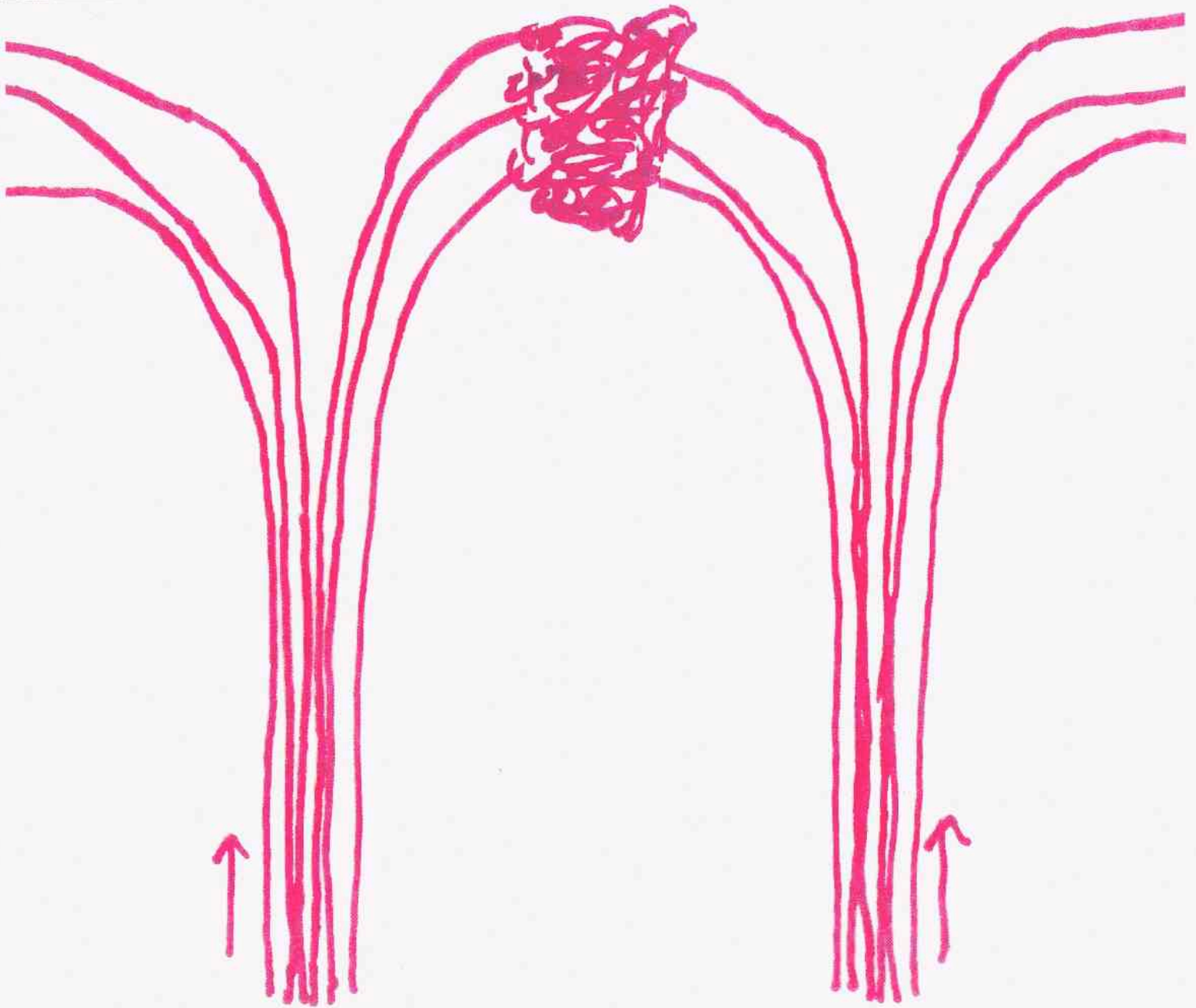

INFALL STOPS RADIATIVE COOLING AND DOWNWARD FLOW BUT NOT UPWARD FLOW

**EXCESS INFALL BLOCKS CONVECTION**

**ENVELOPE AND ATMOSPHERE OUTSIDE THE MAGNETIC CYLINDER PUFFS UP, DECOUPLES, AND SPINS UP TO INCREASE THE MOMENT OF INERTIA**

**COMBINES WITH INFALL GAS THAT IS ALSO MOVING OUTWARD FROM EQUATOR**

**INNER CYLINDER IS A DWARF AND OBLATE PART IS A SUBGIANT**

**THE MATERIAL CAN NEVER GO BACK**

**SINGLE EVENT OR MULTIPLE EVENTS**

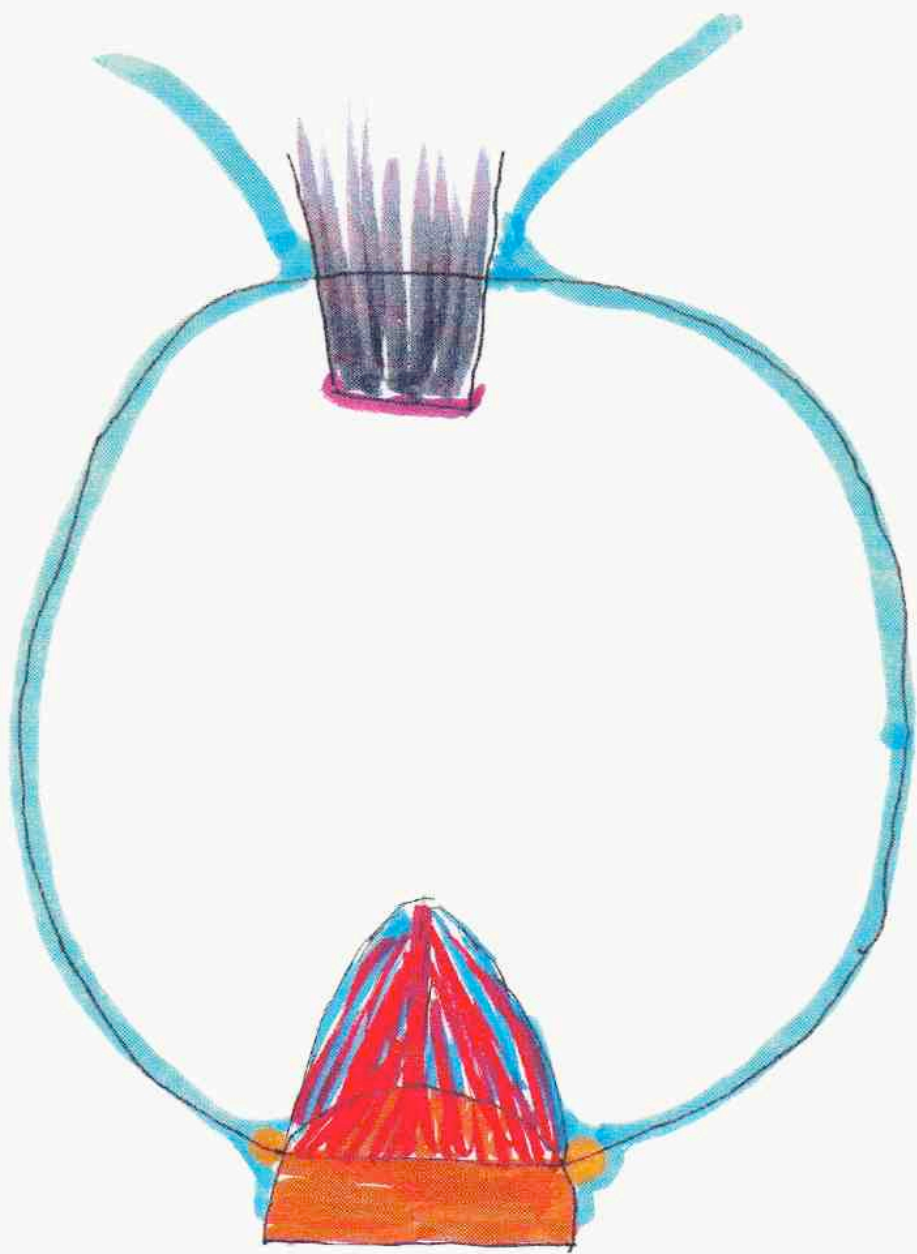

**THE PROTOSTAR, THE MAGNETIC FIELD, AND THE DISK CONTINUE TO EVOLVE AS BEFORE FOR PLANET FORMATION**

**THE COUPLING BETWEEN THE STAR AND DISK WEAKENS AND THE FIELD LINES WRAP**

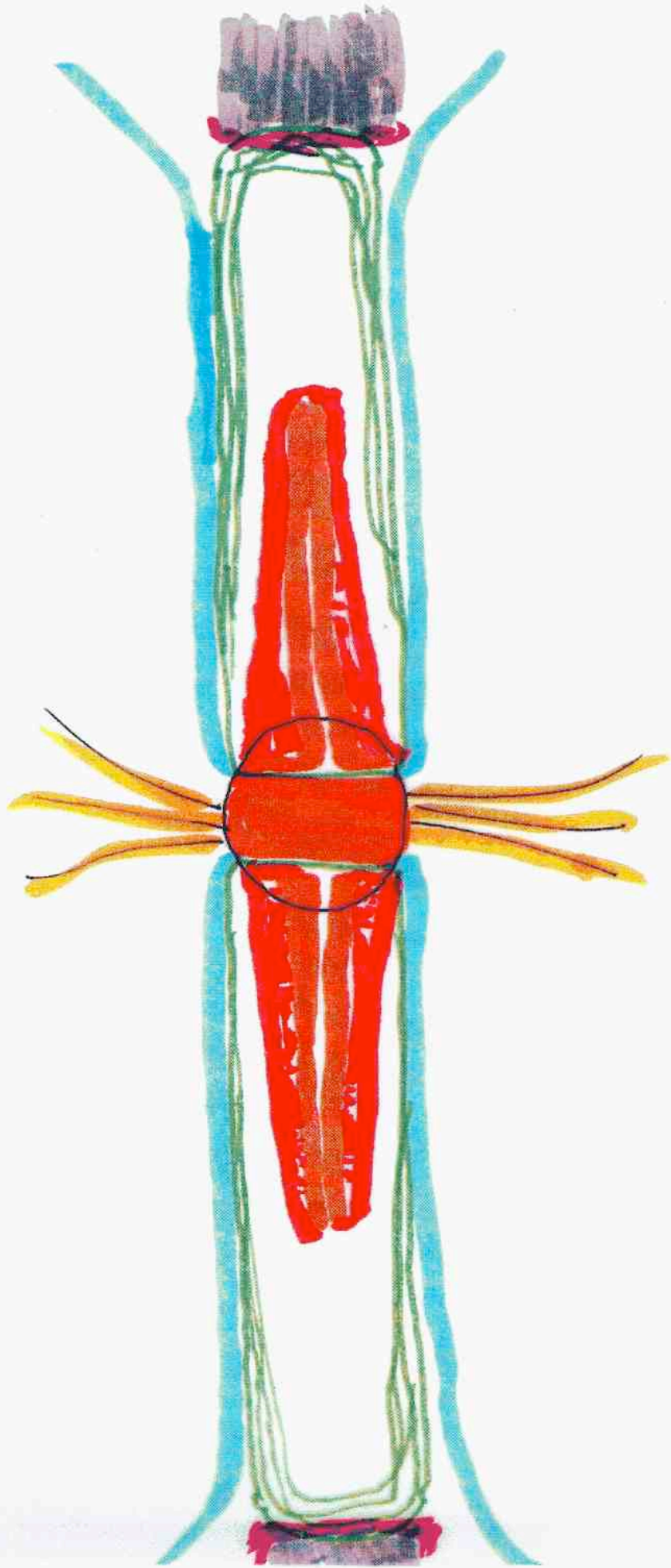

not to scale

64 / 65

IN PLANET FORMATION, FIELD LINES PULL OUT OF THE DISK AND FORM A TORUS FILLED WITH LOW DENSITY IONS THAT COLLAPSE INTO A SUPER-CME

IN CRYPTOPLANET FORMATION, THE FIELD LINES PULL OUT OF THE DISK AND FORM A TORUS FILLED WITH DENSE GAS

INITIALLY THE TOROIDAL MAGNETIC FIELD ROTATES MORE SLOWLY THAN THE GAS INSIDE WHICH HAS HIGH ANGULAR MOMENTUM

THE WHOLE APPARATUS MUST BE HIGHLY UNSTABLE. WHEN A THIN SPOT FORMS FROM OSCILLATIONS OR WHATEVER, THE MAGNETIC FIELD LINES CAN RECONNECT

PART OF THE FIELD FALLS BACK TO MAKE TO MAKE A SUPER-CME
PART FORMS A FREE-SPINNING TOROIDAL RING, A TOKAMAK

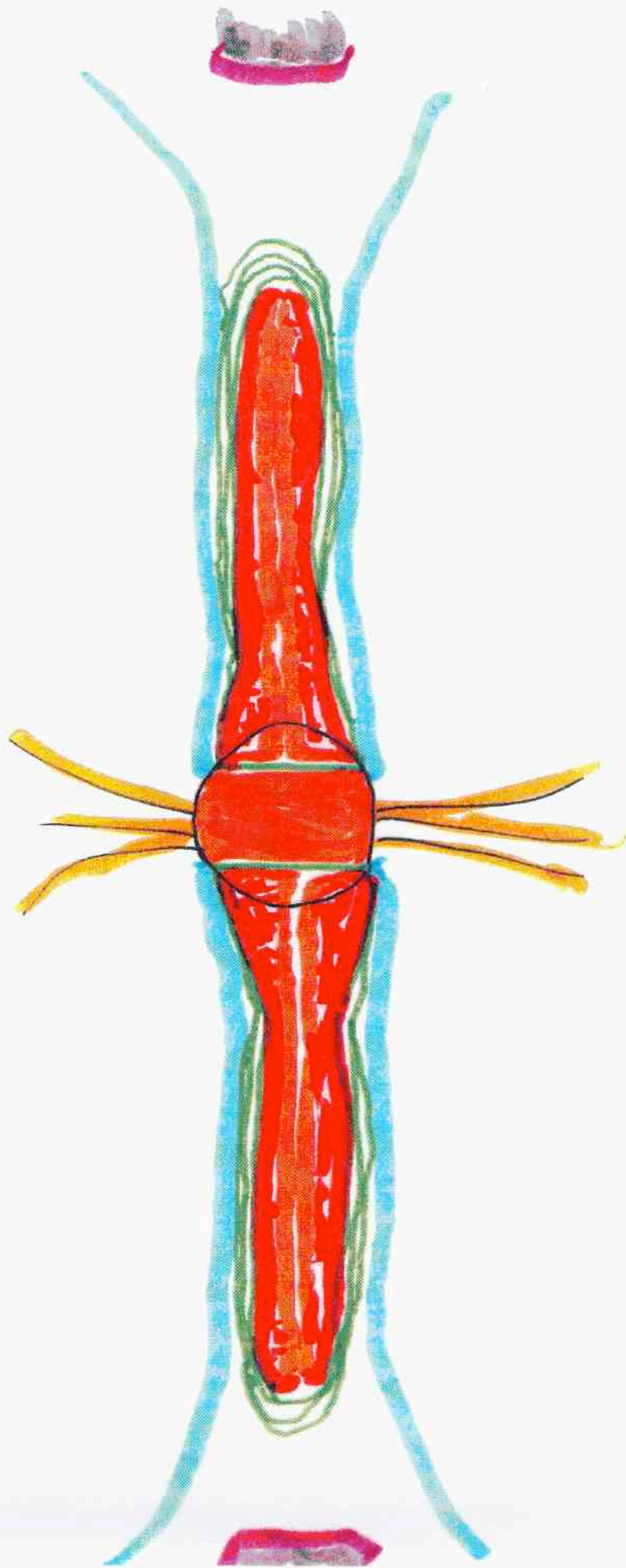

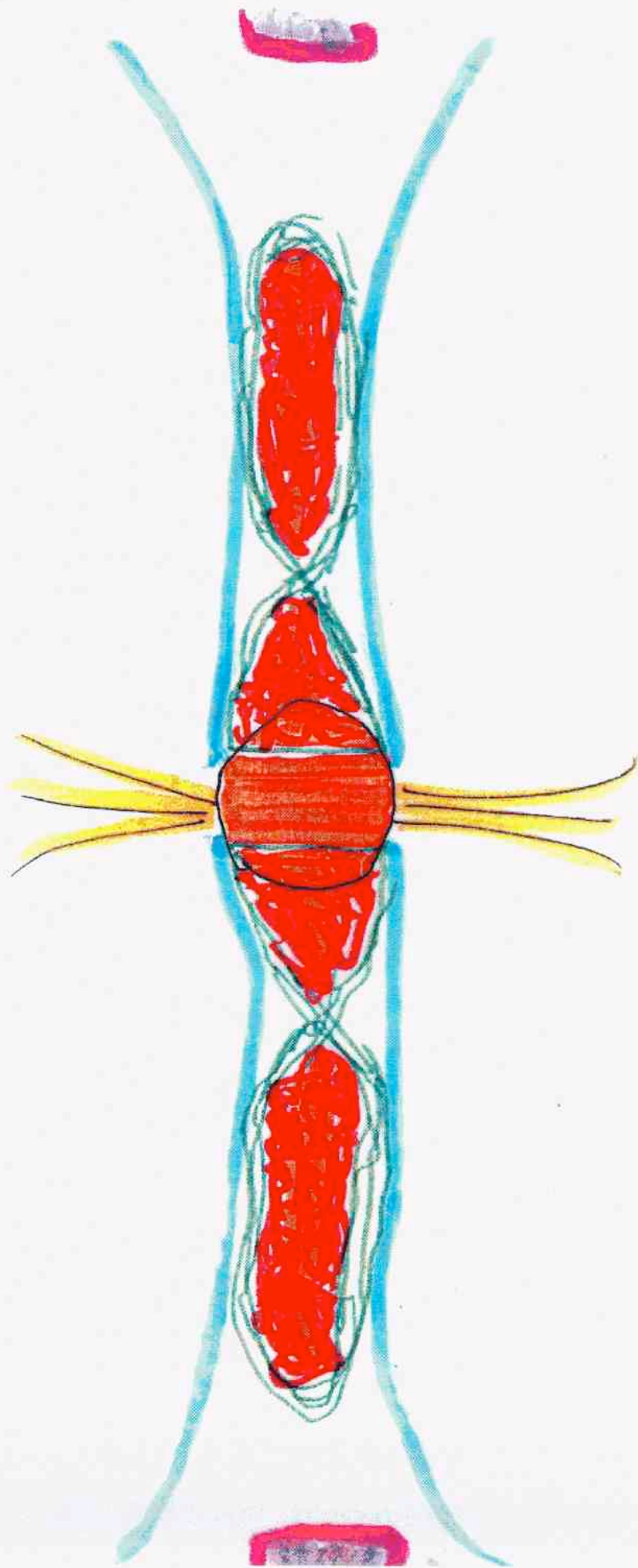

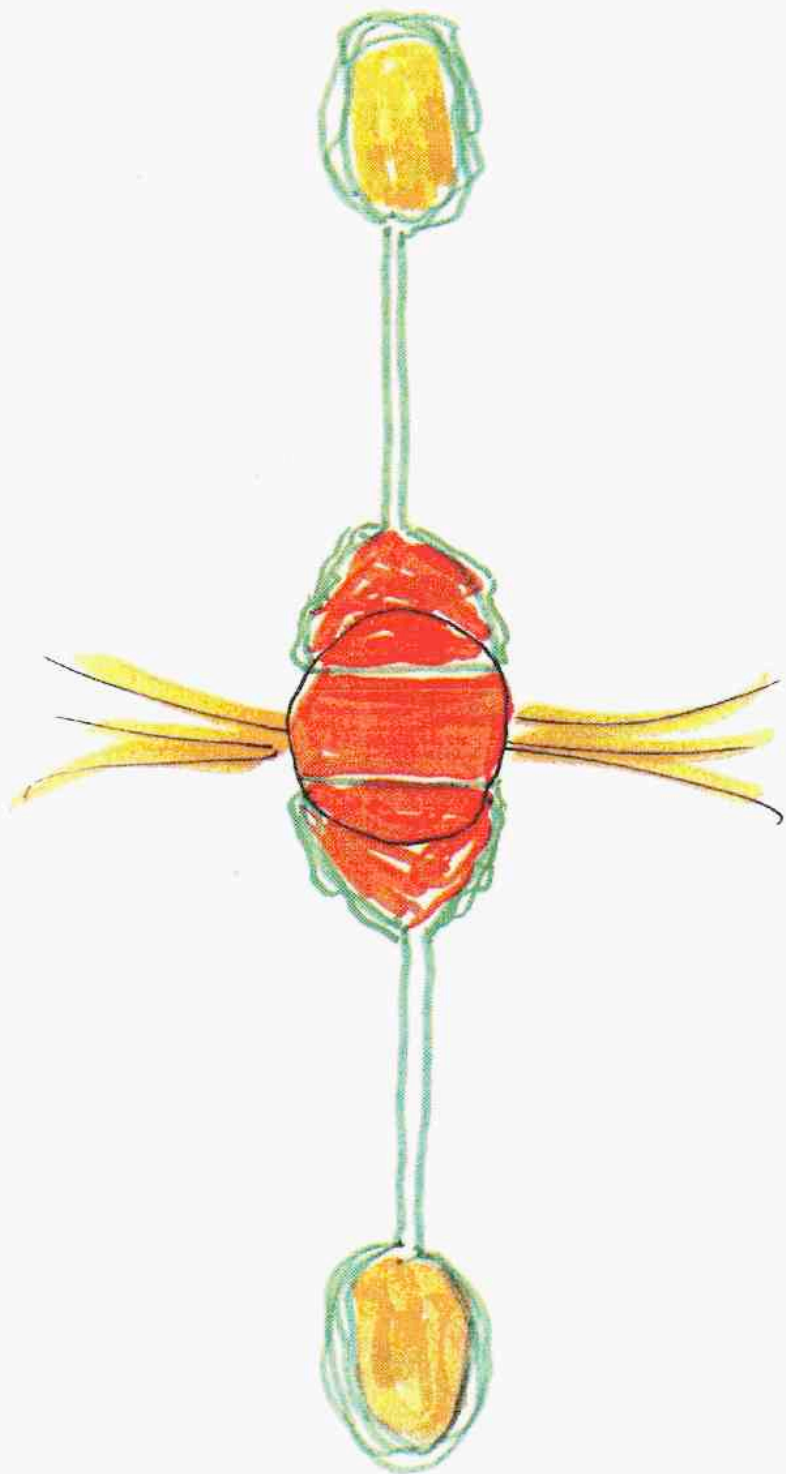

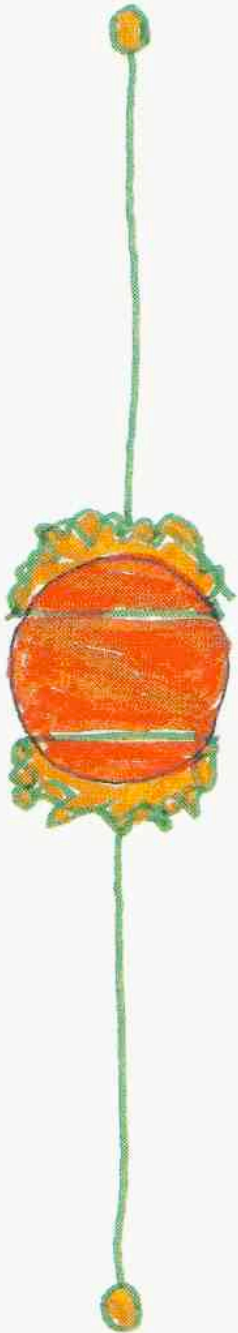

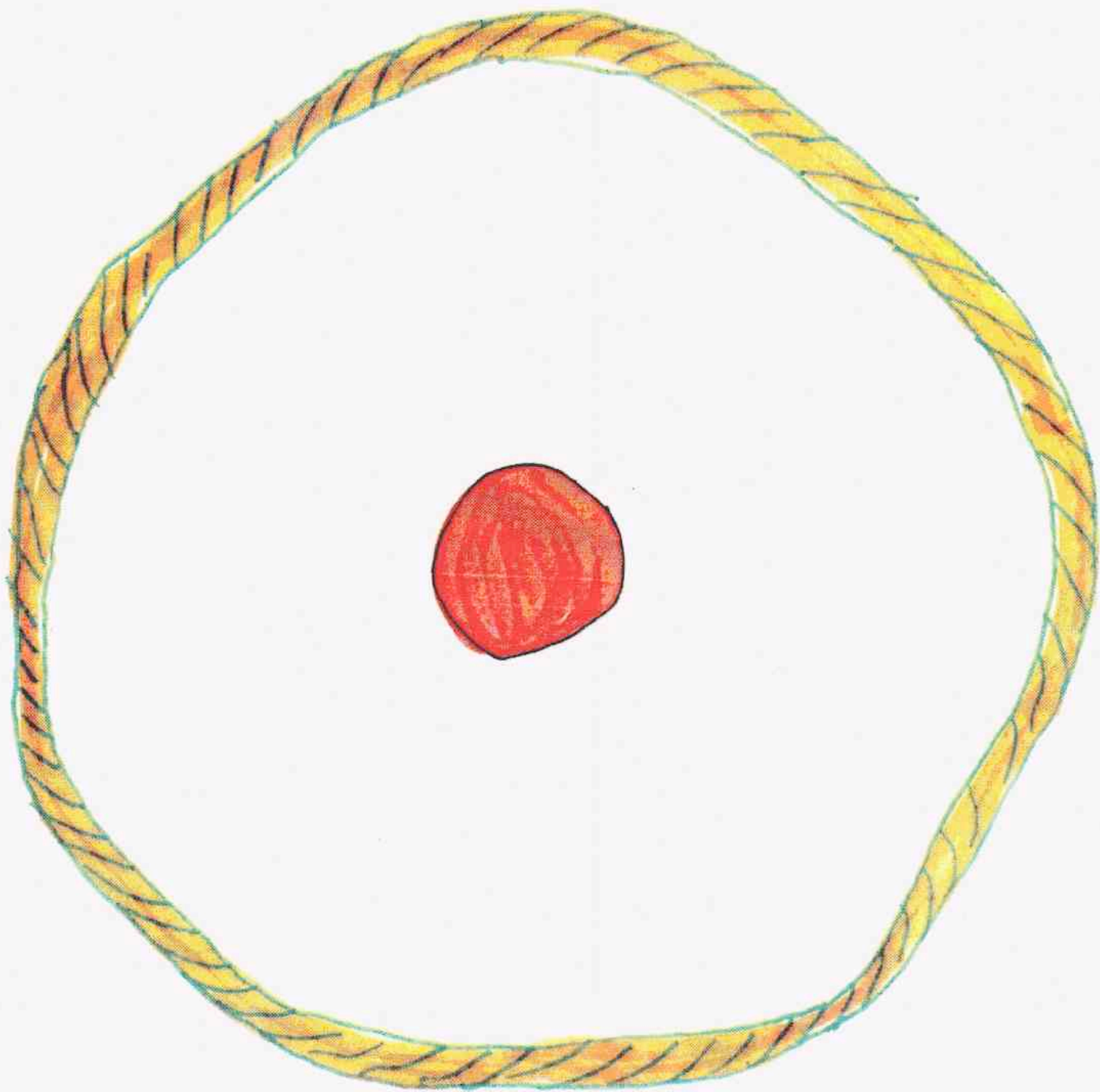

**RING IS HOT AND DENSE AND HAS A LARGE SURFACE AREA**

**RING AND SUPER-CME ARE VERY LUMINOUS. SUPER-CME PRODUCES STRONG PARTICLE OUTBURSTS**

**RADIATION AND PARTICLES BLOW AWAY ALL INFALLING GAS OUT TO ~ SNOW LINE AND PUSH INWARD ON BOTH SURFACES OF THE DISK SO THAT DUST AGGLOMERATES. VAPORIZES AT SURFACE**
**SINCE RING IS EQUITORIAL, IT MAY DESTROY MORE OF THE INNER DISK AND REDUCE INNER PLANET FORMATION**

**RING IS UNSTABLE AND RECONNECTS IN SECTIONS**

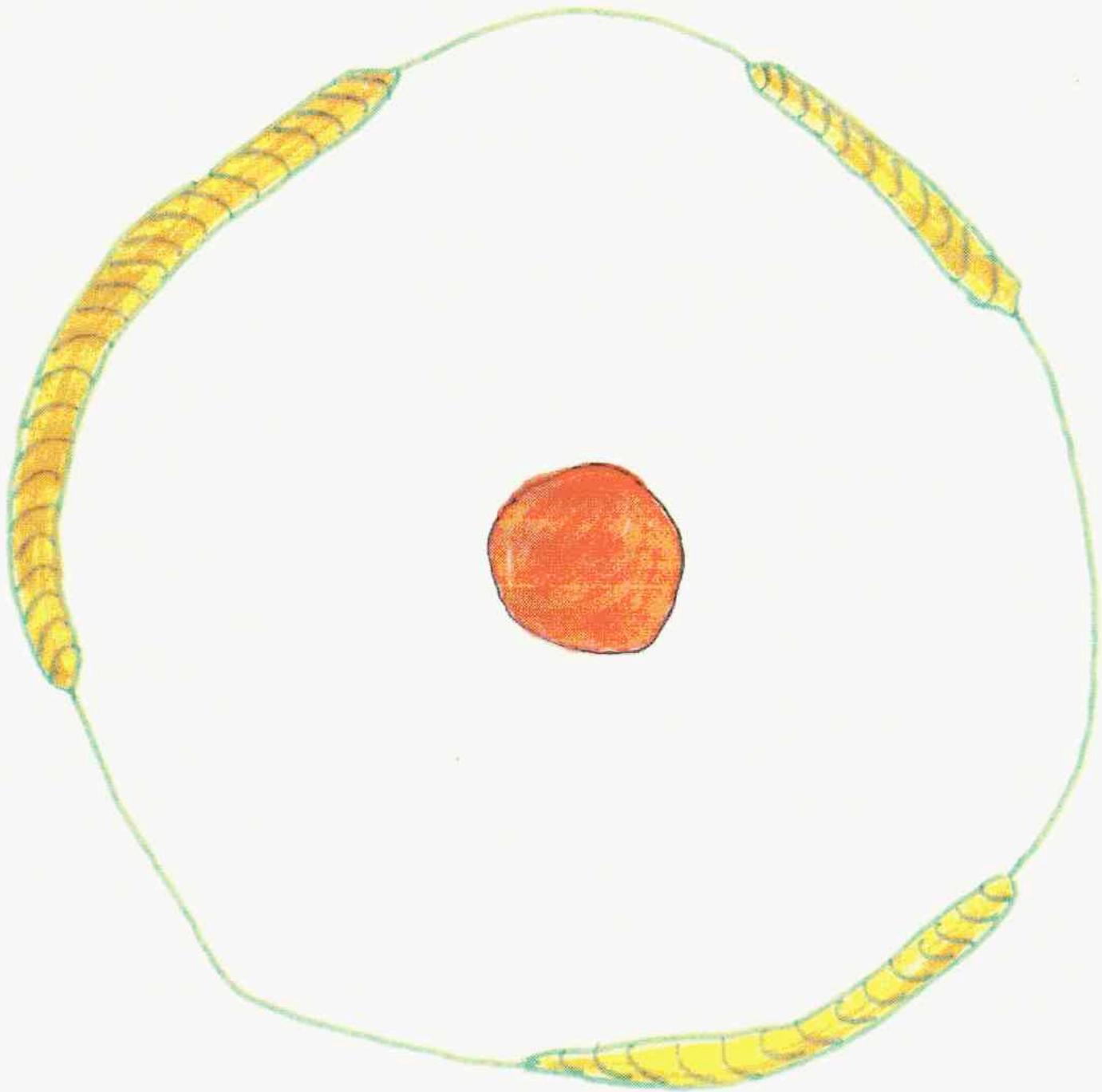

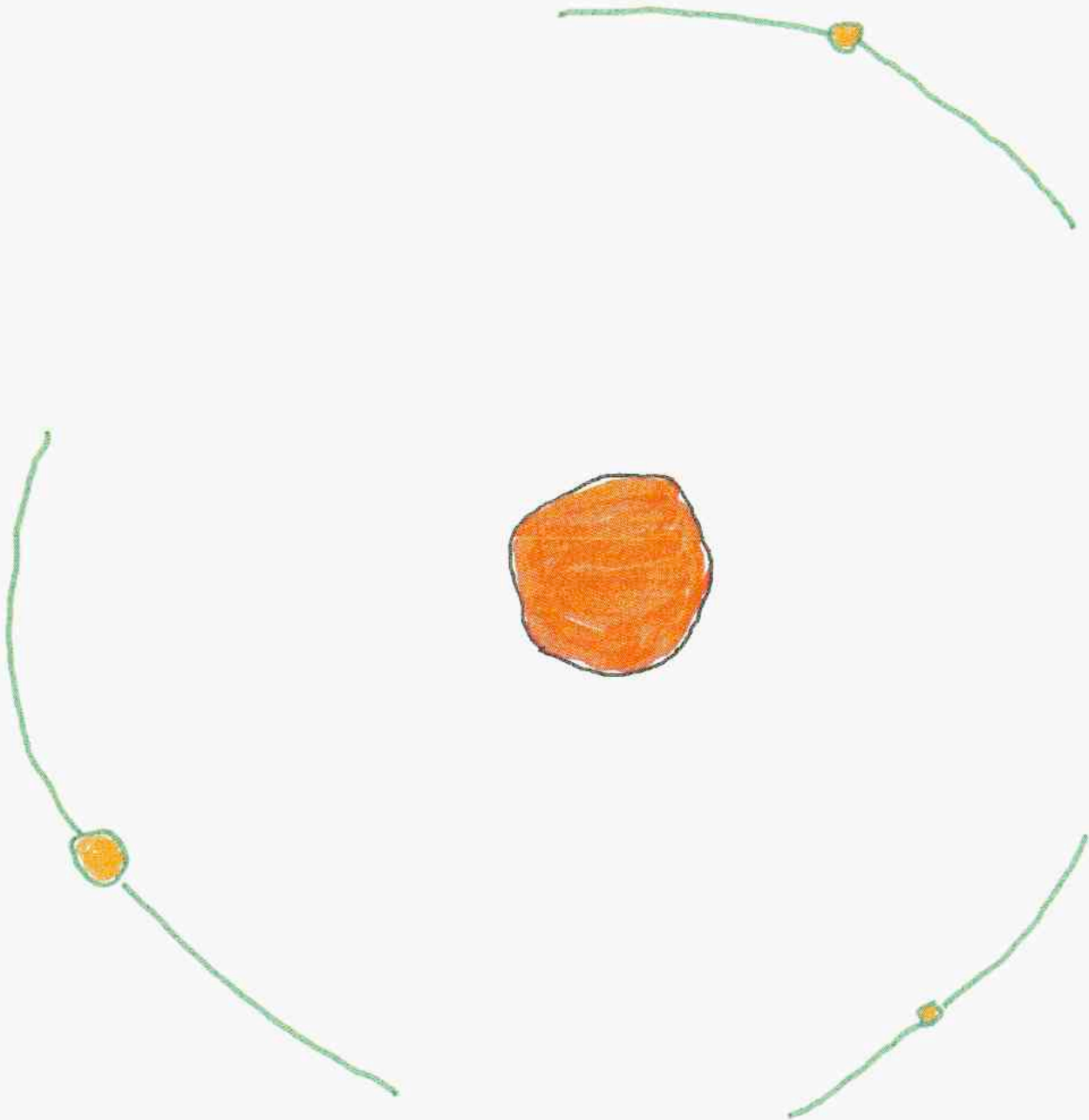

**HOT DENSE PLASMA BLOBS ARE ENCAPSULATED IN SPHERICAL MAGNETIC BOTTLES, SPHEROMAKS.**

**BLOBS ARE LARGE ENOUGH AND DENSE ENOUGH TO BE GRAVITATIONALLY BOUND WHEN THEY COOL, CRYPTOPLANETS**

**MAGNETIC BOTTLES SHIELD THE CRYPTOPLANETS FROM ABLATION BY THE STELLAR WIND**

**THE NUMBER OF CRYPTOPLANETS IS RANDOM, UP TO A DOZEN**

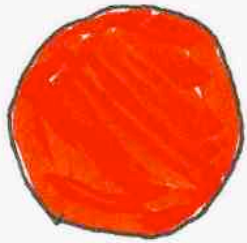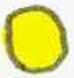

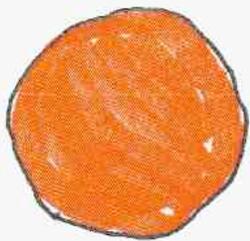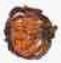

**SOME OF THE CRYPTOPLANETS ARE
    EJECTED FROM THE SYSTEM,
OR FALL INTO THE STAR,
OR EVAPORATE,
OR MERGE,
OR CONTINUE TO ORBIT.**

**CRYPTOPLANETS CAN ACCRETE NEUTRAL GAS**

**MASS CAN VARY UP TO BROWN DWARF MASS**

**BROWN DWARFS CAN BE CRYPTOPLANETS,
    FREE OR BOUND**

# SPHEROMAKS

**A SPHEROMAK IS A SELF-GRAVITATING SPHERE OF HOT IONS OF MOSTLY H AND HE SURROUNDED BY A SPHERICALLY WOUND MAGNETIC FIELD THAT PREVENTS IONS FROM ESCAPING OR ENTERING**

**SPHEROMAKS DO NOT HAVE EXTERNAL MAGNETIC FIELDS**
**DO NOT ROTATE MUCH**
**LARGE ORBITAL ANGULAR MOMENTUM**
**ABUNDANCES UNKNOWN, NOT FROM DUST**
**PROBABLY SUB-SOLAR LIKE YOUNG A-STAR**
    **VEGA, LAMBDA BOO**

**COOLS RAPIDLY BY RADIATION AND CONTRACTS**
**PRESSURE AND DENSITY KEEP INCREASING**

**FORMS HEH+, H2, H2+, H3, H3+, HYDRIDES CARBIDES, NITRIDES, OXIDES, AND THEIR IONS**

# CRYPTOPLANET CHARACTERISTICS

**SPHEROMAKS EVENTUALLY COOL AND BECOME CRYPTOPLANETS**
**COOL FOR A CRYPTOPLANET IS STILL > 1000K.**

**FORM TYPICAL PLANETARY MOLECULES BUT WITH NON-PLANETARY ABUNDANCES**

**OPACITY IS NOT PLANETARY**

**INTERIOR MODELS ARE WRONG**
**CORE NOT AGGLOMERATED BUT LOW DENSITY**

**SURFACE MODELS ARE WRONG**
**PREDICTED ENERGY DISTRIBUTION IS WRONG**
**CANNOT ASSUME SED IN INTERPRETING SPITZER OBSERVATIONS**

**CRYPTOPLANETS ARE A GENERAL PHENOMENON OF POP I UNARY LATE-TYPE STAR FORMATION**

**THEY INDICATE THAT SOMETHING WENT WRONG WITH THE FORMATION PROCESS SO THAT THERE WAS TOO MUCH INFALL**

**THIS MUST HAPPEN MORE THAN 1% OF THE TIME**

**THE CHANCE THAT ONE OR MORE CRYPTOPLANETS WILL CONTINUE TO ORBIT THE STAR IS RANDOM**

**THE MASSES OF CRYPTOPLANETS ARE RANDOM**

**CRYPTOPLANETS MAKE IT DIFFICULT
TO DETECT PLANETS**
    MODULATE RADIAL VELOCITIES
    BRIGHTER THAN PLANETS

**IN ECCENTRIC ORBITS THAY MAY SWALLOW
INNER PLANETS**

<span style="color:red">**IF THE PURPOSE OF SEARCHING FOR PLANETS
IS SEARCHING FOR LIFE, CRYPTOPLANETS
ARE NOT GOOD CANDIDATES**</span>

**WHEN THERE IS TOO MUCH INFALL, THE INFALL FLOWS OUTWARD AT THE EQUATOR AND HITS THE INNER WALL OF THE DISK.**

**THERE IS AN IMMEDIATE CATASTROPHE AND PROBABLY MATTER IS FLUNG OUTWARD.
FU ORI STARS?**

**THE BIRTH OF AN FGK STAR IS DEFINED AS THE PULLOUT OF THE DIPOLE FIELD FROM THE DISK SO THAT INFALL STOPS AND THE PROTOSTAR STOPS GROWING.**

**THE BIRTHRATE FOR FGK STARS IN OUR GALAXY IS BETWEEN 1/DECADE AND 1/CENTURY.**

**IT IS A ONCE IN A LIFETIME EXPERIENCE.**

**THE FORMATION OF A STAR WITH CRYPTOPLANETS HAPPENS ONCE IN A MILLENNIUM**

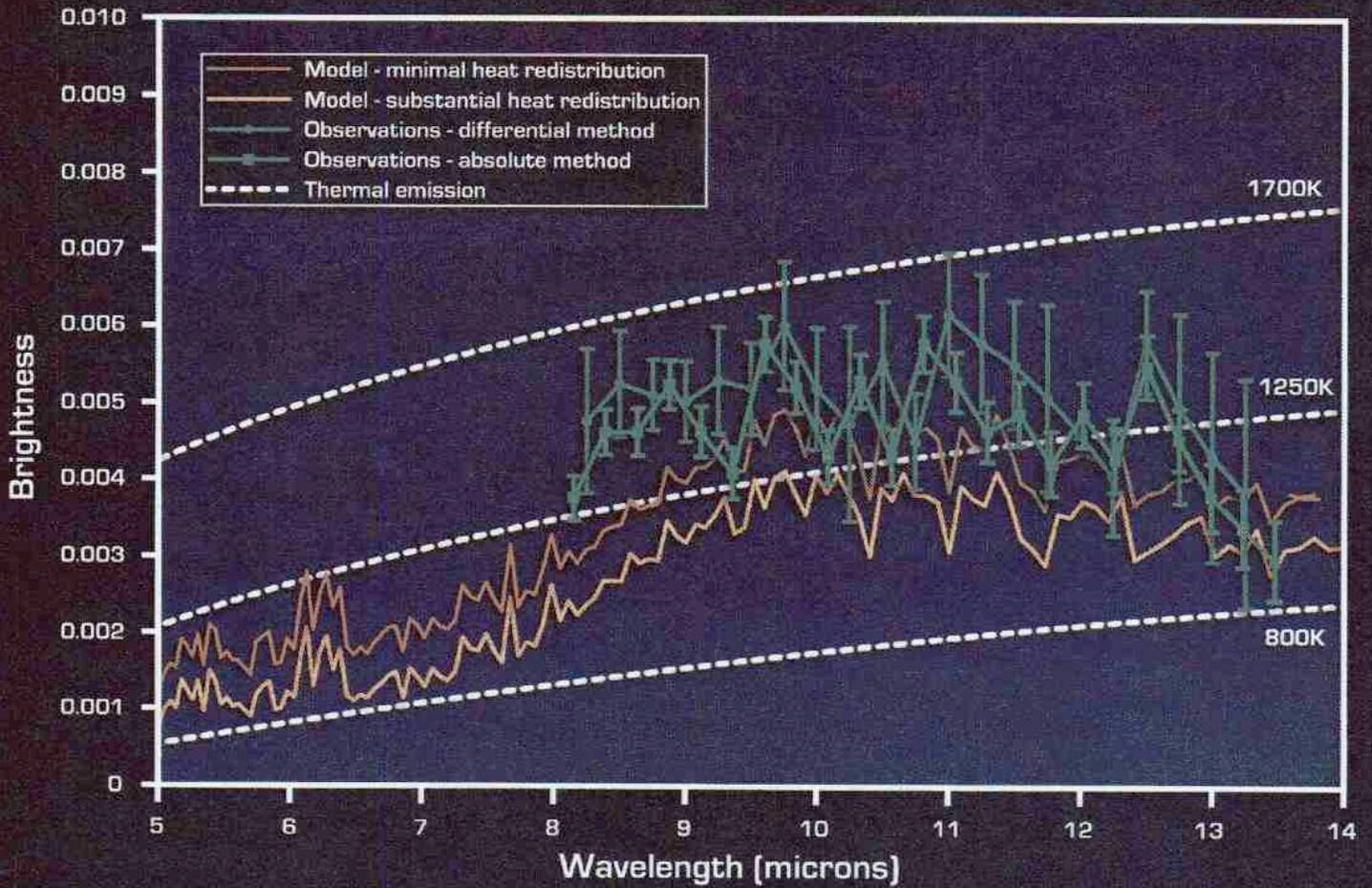

Infrared Spectrum of HD 209458b
NASA / JPL-Caltech / M. R. Swain (JPL/Caltech)
Spitzer Space Telescope • IRS
ssc2007-04b

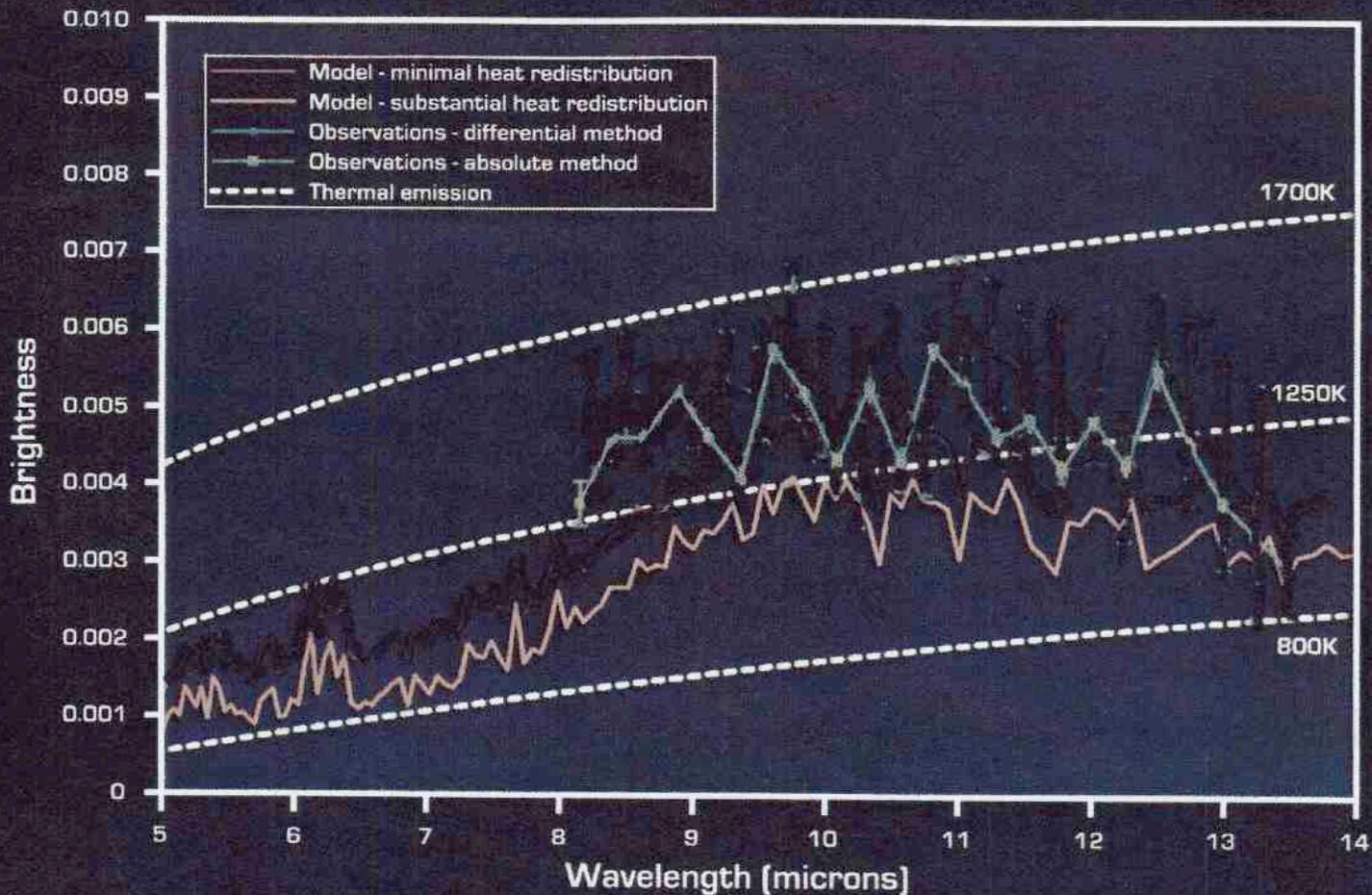

Infrared Spectrum of HD 209458b — Spitzer Space Telescope • IRS
NASA / JPL-Caltech / M. R. Swain (JPL/Caltech)
ssc2007-04b

DE-OBFUSCATED SPECTRA EITHER ANTI-CORRELATED OR NOISE

A3